\documentclass[12pt]{article}
\usepackage{textcomp}
\usepackage{newtxtext,newtxmath}
\usepackage{graphicx}
\usepackage[letterpaper,margin=1.5cm]{geometry}
\usepackage{caption}
\renewenvironment{abstract}
	{\quotation}
	{\endquotation}
\date{}

\makeatletter
\renewcommand{\fnum@figure}{\textbf{Figure \thefigure}}
\renewcommand{\fnum@table}{\textbf{Table \thetable}}
\makeatother
\usepackage{scicite}
\usepackage{url}
\usepackage{xcolor}
\usepackage{soul}
\usepackage{textgreek}
\usepackage{siunitx}

\newcommand{\Hx}{$H_\mathrm{x}$}
\newcommand{\Hy}{$H_\mathrm{y}$}
\newcommand{\Hdmi}{$H_{\mathrm{DMI}}$}
\newcommand{\muHx}{$\mu_0 H_\mathrm{x}$}
\newcommand{\muHy}{$\mu_0 H_\mathrm{y}$}
\newcommand{\tRF}{$t_\mathrm{RF}$}
\newcommand{\pulseT}{$t_\mathrm{p}$}
\newcommand{\MAcm}{MA/cm²}
\newcommand{\PowOutNew}{$P_\mathrm{out}$}
\newcommand{\PowInNew}{$P_\mathrm{in}$}
\newcommand{\jp}{$j_\mathrm{p}$}
\newcommand{\jRF}{$j_\mathrm{RF}$}
\newcommand{\jsw}{$j_\mathrm{sw}$}
\newcommand{\MuSec}{$\mu$s} 
\newcommand{\MuMeter}{\textmu m} 
\newcommand{\Xdisp}{$d_\mathrm{p}$}
\newcommand{\mx}{$m_\mathrm{x}$}
\newcommand{\mz}{$m_\mathrm{z}$}
\newcommand{\Fig}{Fig.}

\newcommand{\dRF}{$d_\mathrm{RF}$}
\newcommand{\dtotal}{$d_\mathrm{total}$}
\newcommand{\DeltaT}{$\Delta t$}

\def\scititle{
    Directed Domain Wall Motion \\ by Radio-Frequency Currents
}
\title{\bfseries \boldmath \scititle}

\author{
	Lukas Fischer$^{1}$, 
	Rouven Dreyer$^{2,3}$, 
	Simon Arnold$^{1}$,
    Jae-Chun Jeon$^{1}$, \\
	Georg Woltersdorf$^{1,2,3}$, 
	Stuart S. P. Parkin$^{1\ast}$\\
	\small$^{1}$Max Planck Institute of Microstructure Physics, Weinberg 2, 06120 Halle (Saale), Germany.\\
	\small$^{2}$Department of Physics, Martin Luther University Halle-Wittenberg, \\ 
	\small Von-Danckelmann-Platz 3, 06120 Halle, Germany. \\
    \small$^{3}$Cluster of Excellence - Center for Chiral Electronics, Halle-Berlin-Regensburg, Germany. \\
	\small$^\ast$Corresponding author. E-mail: stuart.parkin@mpi-halle.mpg.de
}

\begin{document} 
\maketitle

\begin{abstract} \bfseries \boldmath \noindent
Magnetic domain walls (DWs) are central to emerging spintronic technologies for memory, logic and neuromorphic computing, where their controlled displacement is a key functionality. While current-induced DW motion driven by spin-orbit torques is well established under direct currents, it remains unclear how radio-frequency (RF) currents couple to DWs and whether they can generate directed motion under a symmetric periodic drive. Here we show that RF currents drive DW oscillations, which can be rectified into directed motion through symmetry breaking. Using time-resolved and static magneto-optical Kerr microscopy, we demonstrate that RF spin-orbit torques efficiently excite low-frequency DW oscillations well below the ferromagnetic resonance frequency. Whereas these oscillations remain symmetric in the absence of perturbations, symmetry breaking by a transverse magnetic field produces self-propulsion, while symmetry breaking by an initial current pulse gives rise to sustained DW motion. We further show that in initially uniform magnetic states, RF-driven expansion of a pulse-generated edge domain ultimately results in efficient RF-assisted magnetic switching. Together, these results establish a general framework for understanding how RF-driven DW oscillations are rectified into directed motion, providing new opportunities for their high-frequency control.
\end{abstract}
\newpage

\noindent
Current-induced spin-orbit torques (SOTs) provide an efficient mechanism to drive the motion of chiral Néel domain walls (DWs) in heavy-metal/ferromagnet heterostructures \cite{2011Miron, 2013Emori, 2013Ryu, 2019Velez}, enabling spintronic device concepts such as racetrack memory, DW logic and neuromorphic computing \cite{2008ParkinRTM, 2020Luo, 2021SOT, 2022RevDWM, 2023RevDWM, 2024Jeon}.
These torques can propel DWs at high velocities and have been extensively studied under direct-current excitation, where the dynamics are governed by steady or pulsed driving forces. By contrast, radio-frequency (RF) currents generate oscillatory SOTs and thereby enable fundamentally different regimes of DW dynamics, yet their interaction with DWs remains largely unexplored.

In general, RF excitations are widely used to probe and manipulate magnetic systems, most prominently through ferromagnetic resonance (FMR) and spin-wave excitations. FMR describes the coherent precession of a uniform magnetization \cite{1993HeinrichCochranFMR}, whereas propagating spin waves can interact with DWs through the transfer of linear and angular momentum \cite{2021ReviewSchultheissSWDW}. These interactions can lead to DW displacement as well as spin-wave reflection and phase shifts \cite{2019HanLiuScience, 2023FanRoss, 2011YanWang, 2004HertelKirschner}.
However, these phenomena differ fundamentally from the direct excitation of the internal dynamics of a DW. Unlike the uniform magnetization or propagating spin waves, DWs constitute non-uniform magnetic textures with their own internal degrees of freedom and have been shown to respond at frequencies well below the uniform FMR mode \cite{2016Wagner}. In particular, it remains unclear how RF spin torques couple directly to the intrinsic dynamics of a DW and how such excitation can be converted into controlled, directed DW motion under a symmetric drive through rectification.

These observations point to the absence of a unified understanding of RF-driven DW dynamics. In particular, it remains unclear how RF spin torques couple to the internal degrees of freedom of a DW and how they influence its average position under a purely harmonic drive without a static bias. More generally, the mechanisms by which periodic spin torques are converted into directed motion remain unknown, limiting the ability to systematically control DW dynamics under RF excitation.

Here, we aim to close this gap by studying the interaction of RF currents with magnetic DWs in heavy-metal/ferromagnet heterostructures. Using a variant of time-resolved magneto-optical Kerr microscopy together with static Kerr imaging, we first demonstrate that RF currents selectively excite DW oscillations at frequencies well below the uniform FMR mode. This excitation leads to oscillatory DW motion with a stable average position under symmetric driving. When this symmetry is broken, either by an external magnetic field or by a controlled initial current pulse, the oscillatory motion is converted into a net displacement, enabling directed propagation under a purely alternating drive. Furthermore, we identify a regime in which DWs continue to move following an initial excitation in the presence of RF currents, indicating a sustained dynamical response beyond the duration of the driving pulse. Finally, we demonstrate that the same RF-driven dynamics also influence magnetic switching by promoting the expansion of nucleated domains during reversal.

\subsection*{Direct Excitation of Domain Wall Oscillation}
\begin{figure}
    \centering
    \includegraphics[width=\linewidth]{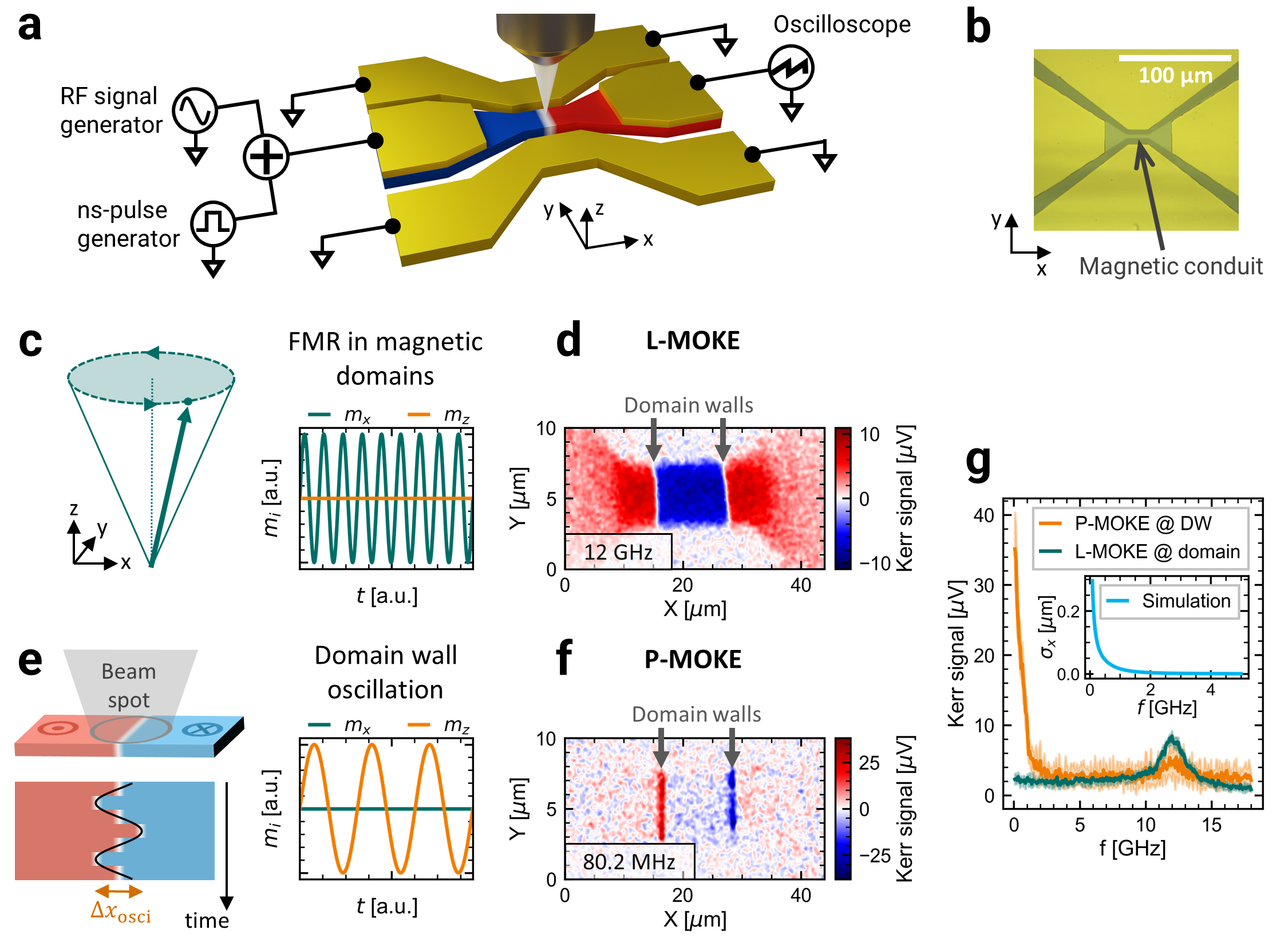}
    \caption{\textbf{Selective excitation of domain wall oscillation and FMR by RF currents.} 
    (\textbf{a}) Schematic of device design, electrical circuitry and optical detection. 
    (\textbf{b}) Microscopy image of device with a straight racetrack with \mbox{$3 \times 20$ \unit{\micro\metre\squared}} dimensions at the center. 
    (\textbf{c}) Schematic of FMR in the uniform domains leading to a periodic \mx{}. 
    (\textbf{d}) Experimental SNS-MOKE map of the racetrack using L-MOKE at the FMR frequency. The measurement is only sensitive to the uniform domains but not to the DWs. For all measurements, an RF power of \PowInNew{} $=10$ dBm is applied to the device input and two magnetic DWs are placed in the racetrack (from left to right: up-down and down-up). 
    (\textbf{e}) Schematic of DW oscillation leading to a periodic \mz{} variation. 
    (\textbf{f}) Experimental SNS-MOKE map using P-MOKE at a lower frequency of $f=80.2$ MHz. This measurement is only sensitive to the DWs. 
    (\textbf{g}) Comparison of the frequency dependence of FMR in the domains (green) and the DW oscillation (orange). The standard deviation from multiple measurements at each frequency is indicated by a shaded area. The inset displays the standard deviation of the DW oscillation, obtained from numerical simulations using the $q-\psi-\chi$ model using an RF current density of $100$ \MAcm{}. All experiments and simulations were performed in zero magnetic field.
    }
    \label{fig_main: Intro}
\end{figure}

To investigate how RF spin torques couple to magnetic DWs, we use magneto-optical techniques to study chiral Néel DWs in perpendicularly magnetized ferromagnetic racetrack devices (see Methods). The film stack exhibits Dzyaloshinskii-Moriya interaction (DMI) and also allows for a current-induced SOT. To conduct the experiment, we first initialize one or more DWs in the racetrack and then inject an RF current directly through the magnetic racetrack (Figs. \ref{fig_main: Intro}a, b).
The RF current generates periodic spin torques that act on the magnetic system, resulting in two effects. When acting on the uniform magnetic domains, the RF spin torques can excite FMR. Additionally, since a DW moves under the action of current-induced spin torques (see Methods), the alternating drive will induce an oscillation of the DW. 
We optically probe the magnetization dynamics of the sample using the Super-Nyquist sampling magneto-optical Kerr effect (SNS-MOKE) microscopy. SNS-MOKE provides two complementary measurement configurations that selectively probe either dynamics of the DW (perpendicular, or P-MOKE) or the surrounding domains (longitudinal, or L-MOKE).

L-MOKE is sensitive to the dynamic \mx{}$(t)$ component and therefore allows to detect the FMR in the uniform out-of-plane domains in which the precession cone lies in the $x$-$y$ plane (Fig. \ref{fig_main: Intro}c). In contrast, an oscillation of a Néel-type DW within the beam spot does not lead to any temporal change in \mx{} and thus does not produce a measurable L-MOKE signal (Fig. \ref{fig_main: Intro}e). Indeed, the L-MOKE map in Fig. \ref{fig_main: Intro}d shows strong signal in the uniform domains (blue and red regions) and zero signal at the DWs (white lines). Note that the color jump from red to blue corresponds to a 180° phase shift of \mx{}$(t)$ and is expected due to the opposite equilibrium magnetization on the different sides of the DW.

P-MOKE, on the other hand, is only sensitive to the dynamic \mz{}$(t)$ component. While the uniform FMR mode produces no periodic variation of \mz{}, an oscillating DW within the laser spot gives rise to such a modulation and therefore generates a P-MOKE response (Figs. \ref{fig_main: Intro}c, e). The P-MOKE map in Fig. \ref{fig_main: Intro}f shows a magnetic response only at the positions of the two DWs (red and blue lines) and very low signal in the uniform domains. The phase change between the left and right wall is associated to the fact that the periodic \mz{}$(t)$ variation is 180° phase-shifted for an up-down DW and down-up DW, respectively.

Finally, Fig. \ref{fig_main: Intro}g displays the frequency response of the FMR in the uniform domain (using L-MOKE) and the DW oscillation (using P-MOKE). The FMR response from L-MOKE (green data) is centered at \mbox{$f \approx 12$ GHz} and agrees well with the extrapolated zero-field resonance from conventional FMR measurements on extended magnetic films (Supplementary Fig. \ref{fig-SI: FMR}) and the effective anisotropy field deduced from magnetometry measurements (Supplementary Fig. \ref{fig-SI: VSM}). However, the P-MOKE signal associated with the DW oscillation (orange data) is negligible at the FMR frequency, where only a weak residual signal originating from L-MOKE contributions in the surrounding domains is detected. At lower frequencies, a pronounced P-MOKE response emerges below 2 GHz. Upon further decreasing the RF frequency, the signal amplitude continuously increases, and down to the lowest investigated frequency of 80 MHz, no peaking or saturation of the P-MOKE response is observed. Furthermore, the P-MOKE signal associated with the DW oscillation exhibits a considerably stronger dependence on RF power than the corresponding FMR signal detected by L-MOKE (Supplementary Fig. \ref{fig-SI: SNS power dep}), further highlighting their distinct dynamical response. 

To gain insight into the frequency dependence of the DW oscillation, we performed numerical simulations based on the well-established collective coordinate model $q-\psi-\chi$ \cite{2002Thiaville, 2012Thiaville, 2013Ryu, 2013Boulle} that treats the DW as a one-dimensional object with a position $q$, an in-plane angle $\psi$ and a DW tilt $\chi$ (see Methods). 
We simulated the impact of an RF-current (or RF spin torques) on the DW and confirm that it indeed induces a DW oscillation. From the time traces of this symmetric oscillation, we then extracted the standard deviation $\sigma_\mathrm{x}$ of the DW position, which serves as a measure of the oscillation amplitude. The frequency-dependence of $\sigma_\mathrm{x}$ is shown in the inset of \mbox{Fig. \ref{fig_main: Intro}e} and is in good qualitative agreement with the experimental data. Altogether, these measurements identify a distinct low-frequency oscillation associated with the DW that is spectrally separated from the uniform FMR response. In the following, we exploit this selective low-frequency excitation to investigate how symmetric and symmetry-broken DW oscillations evolve into a directed DW motion.

\subsection*{RF-driven Self-Propulsion of Magnetic Domain Wall}
By symmetry, a DW driven by positive and negative current polarities undergoes equal but opposite displacements \cite{2013Emori, 2013Ryu}, resulting in zero net motion under a purely oscillatory RF drive. This symmetry in the DW's response to positive/negative currents is broken in the presence of a transverse magnetic field \Hy{}. Such a static \Hy{} field, while not  capable of continuously moving the DW by itself, enhances the current-induced DW motion for one current polarity and suppresses it for the opposite polarity (see Methods) \cite{2013Ryu}. This suggests that an RF current with alternating current polarity will induce an asymmetric DW oscillation that is rectified into a directed motion.

\begin{figure}[h!]
    \centering
    \includegraphics[width=\linewidth]{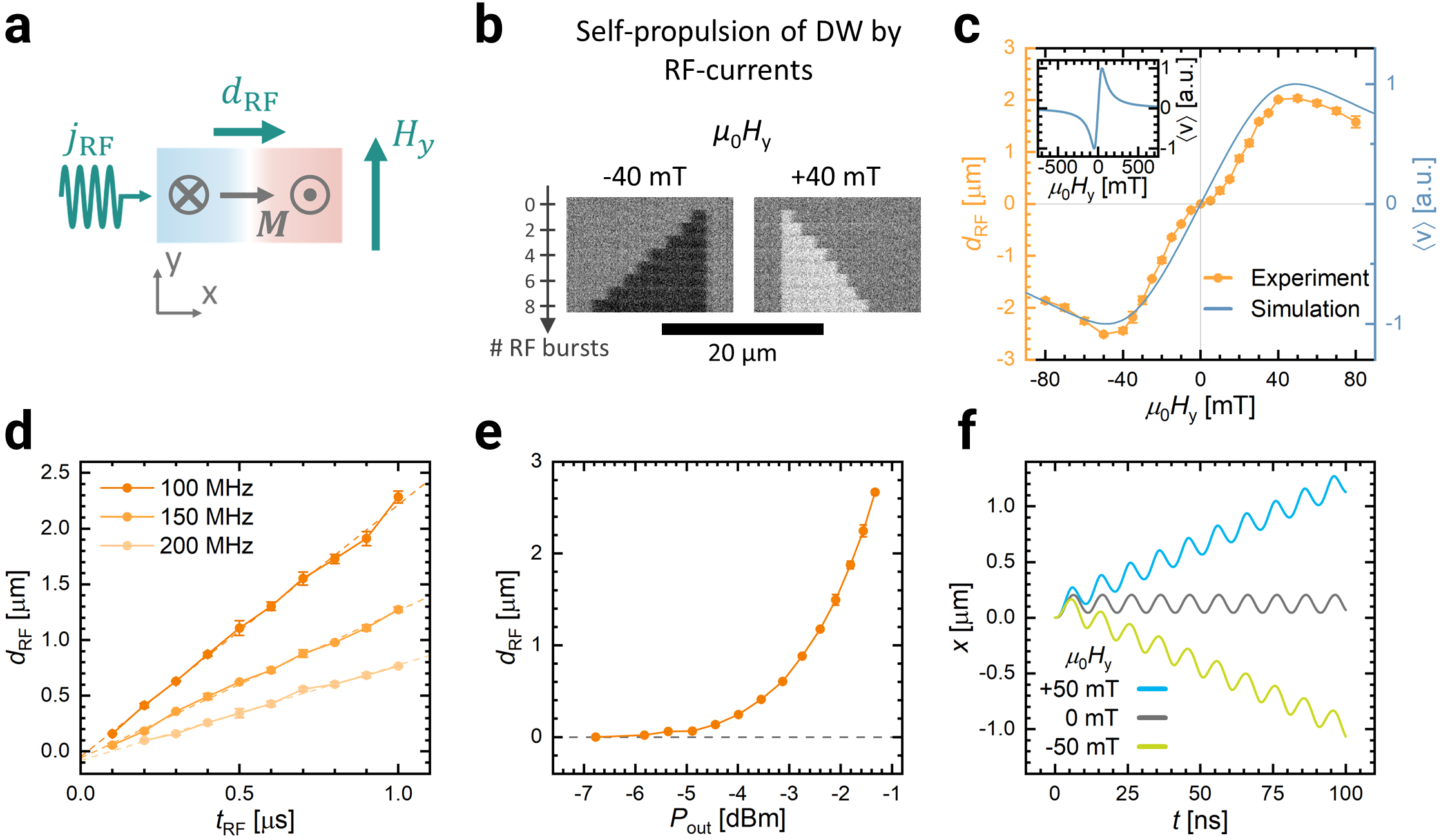}    
    \caption{
    \textbf{Self-propulsion of domain wall by rectified oscillation.} 
    (\textbf{a}) Schematic of DW configuration, RF-current injection and transverse magnetic field \Hy{}. 
    (\textbf{b}) Kerr microscopy images of gradual RF-driven DW motion after application of one RF burst per image line at $f=250$ MHz, \tRF{} $=1$ \MuSec{} and \PowOutNew{} $=-1.9$ dBm. White (black) contrast indicates that the down-up DW is moving to the right (left).
    (\textbf{c}) \Hy{}-field dependence of self-propulsion for $f=100$ MHz, \tRF{} $=0.5$ \MuSec{} and \PowOutNew{} $=-1.84$ dBm. The blue data in plot and inset are the normalized mean DW drift velocities obtained from numerical simulations based on the $q-\psi-\chi$ model. The simulations used $f=100$ MHz and an RF current density \jRF{} $=7$ \MAcm{}.
    (\textbf{d}) Linear increase of DW displacement with RF-duration for \PowOutNew{} $=-3$ dBm and \muHy{} $=40$ mT. 
    (\textbf{e}) Power dependence of self-propulsion for $f=250$ MHz, \tRF{} $=1.0$ \MuSec{} and \muHy{} $=40$ mT. (\textbf{f}) Simulated time-traces of the DW position during application of an RF-current starting with zero phase at $t=0$. The same RF parameters as in (\textbf{c}) are used.
    }
    \label{fig_main: self prop}
\end{figure}

Static Kerr microscopy directly confirms this rectification experimentally. The static imaging of the position of a single down-up DW allows to to detect the DW's displacement (\dRF{}) after the application of RF-current with a burst length \tRF{}, typically in the range from $100$ ns to several \MuSec{} (see Methods). Fig. \ref{fig_main: self prop}b displays the gradual DW motion upon passing RF-current at $f=250$ MHz with \tRF{} $=1$ \MuSec{} in the presence of a transverse field with magnitude \mbox{$|$\muHy{}$|$ $=40$ mT}. In a positive \Hy{}-field, the DW moves along $+x$ during each RF-burst. In contrast, for a negative \Hy{}-field, the direction of motion reverses. The detailed \Hy{}-field dependence of the DW-displacement \dRF{} per RF-burst in Fig. \ref{fig_main: self prop}c confirms that no motion occurs in zero field and that \dRF{} is antisymmetric with respect to \Hy{}, i.e. \dRF{}$(+$\Hy{}$) \approx -$\dRF{}$(-$\Hy{}$)$. Moreover, the displacement reaches a maximum around $|$\muHy{}$| \approx 50$ mT. Control measurements confirm that a residual out-of-plane field component is insufficient to independently drive DW motion (see Methods). These results demonstrate how a transverse magnetic field can be utilized to turn a symmetric DW oscillation in a periodic drive current into unidirectional motion. In the following, we refer to this rectified RF-driven DW motion as \textit{self-propulsion}.

The presence of self-propulsion not only depends on the symmetry-breaking field, but also on the RF-parameters such as burst duration, RF power, and frequency. Fig. \ref{fig_main: self prop}d shows that \dRF{} increases linearly with the RF burst duration. This demonstrates that the rectified motion occurs in a continuous fashion throughout the burst rather than originating from a single displacement event. Moreover, the average propagation rate is largest at the lowest RF frequencies, in excellent agreement with the frequency dependence of the DW oscillation identified in Fig. \ref{fig_main: Intro} (also see Supplement Fig. \ref{fig-SI: self-prop frequency}). The linear dependence corresponds to a constant average propagation velocity of, for instance, $v_{\mathrm{avg}} = 2.26$ m/s for $f = 100$ MHz. The RF-driven DW motion exhibits a clear excitation threshold as revealed by the RF-power dependence in Fig. \ref{fig_main: self prop}e. While at low \PowOutNew{}, no motion is observed, significant DW displacements occur for \PowOutNew{} $\geq -4.5$ dBm. These results show that self-propulsion arises from the continuous rectification of a DW oscillation and becomes observable once the RF excitation exceeds a threshold amplitude. While the experiments establish steady self-propulsion, the microscopic origin of this rectification cannot be inferred from the static images. We therefore employ numerical simulations to understand the mechanism behind these observations.

In particular, we use the previously introduced collective coordinate model to numerically study RF-driven DW oscillation and how an \Hy{}-field rectifies this oscillation into directed motion. The simulated time traces in Fig. \ref{fig_main: self prop}f reveal symmetric oscillations in zero field, whereas applying positive (negative) \Hy{} produces a finite drift in the positive (negative) $x$-direction. The simulated trajectories demonstrate that the oscillation remains periodic but becomes asymmetric with respect to the current polarity. For instance, for \Hy{} $> 0$ the motion during a positive (negative) current polarity is enhanced (suppressed), leading to a net motion along $+x$. This rectification converts the periodic oscillation into a finite drift velocity. Extracting the mean DW velocity for a range of \Hy{}-fields allows to compare the simulations with the experimental data in Fig. \ref{fig_main: self prop}c (blue data, inset) and reveals a good agreement. This confirms that the mechanism for self-propulsion is indeed the rectification of a DW oscillation that is induced by a symmetry-breaking transverse magnetic field \Hy{}.

\subsection*{Current-Triggered Sustained Domain Wall Motion} 
\begin{figure}[hb!]
    \centering
    \includegraphics[width=\linewidth]{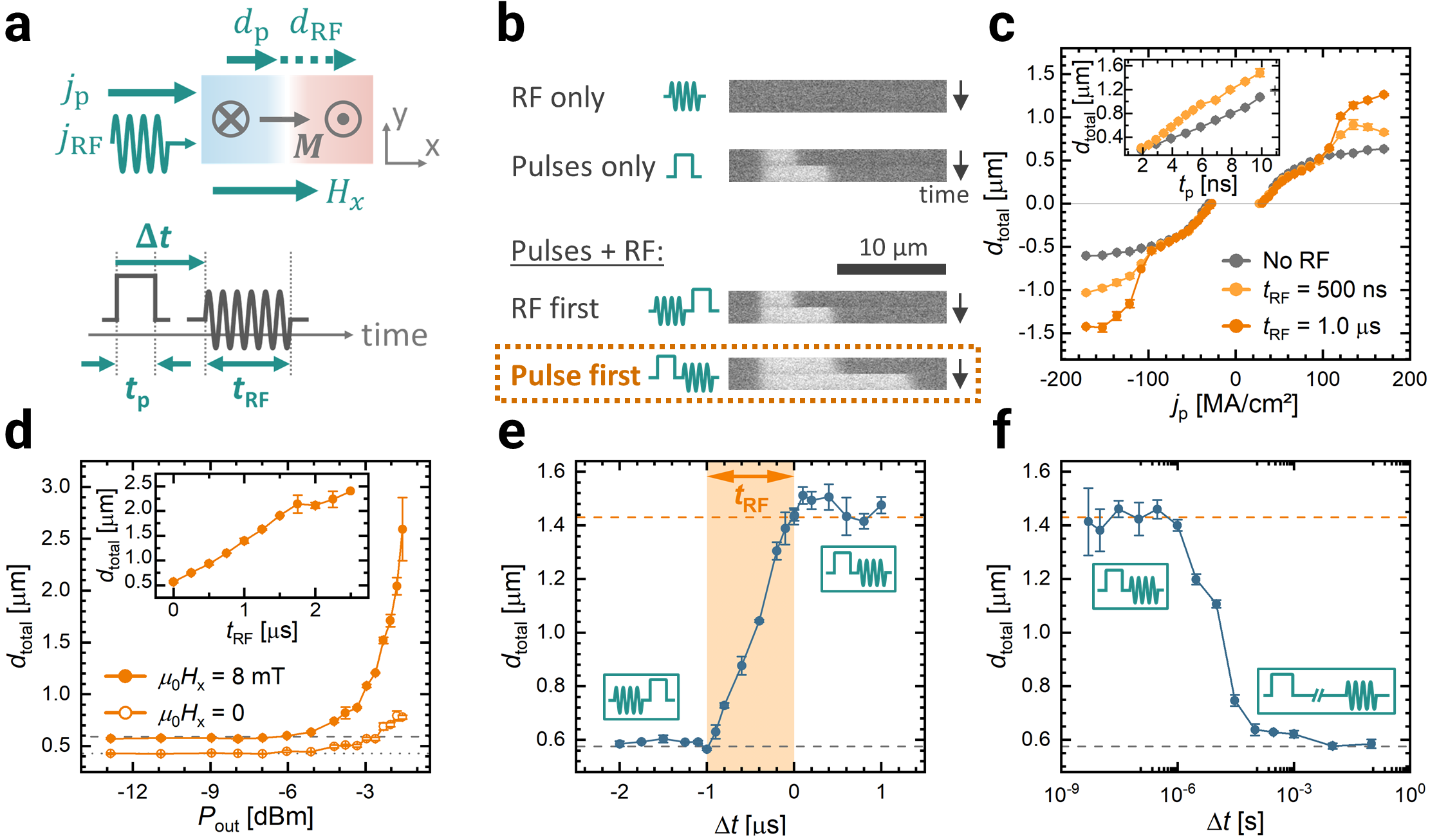}    
    \caption{\textbf{Sustained domain wall motion.} (\textbf{a}) Top: schematic of DW configuration, current injection (both pulse and RF) and longitudinal magnetic field. Bottom: schematic of pulse and RF duration and their relative timing. 
    Unless differently specified, the following parameters were applied for the experimental results: \mbox{\jp{} $=136$ \MAcm{}}, \pulseT{} $=5$ ns and RF-bursts with $f=250$ MHz, \PowOutNew{} $=-3$ dBm, \tRF{} $=1.0$ \MuSec{}. The magnetic field was set to \muHx{} $=8$ mT. 
    (\textbf{b}) Kerr microscopy images of different combinations of pulses and RF-current. For each image line, six pulses and/or RF-bursts were applied. The white contrast indicates that the down-up DW is moving to the right.
    (\textbf{c}) Current density dependence of the DW displacement for different RF conditions. The inset displays the effect of the current pulse width \pulseT{} on the DW motion. The pulses in (\textbf{c-d}) were sent immediately before the RF-burst, i.e. $\Delta t \approx 10$ ns. 
    (\textbf{d}) Dependence of \dtotal{} on the transmitted RF-power using bursts with \tRF{} $=0.5$ \MuSec{}. The inset displays the dependence of the DW motion on the duration of the RF-burst. 
    (\textbf{e-f}) Timing dependence between current pulse and RF-burst for linear (\textbf{e}) and logarithmic (\textbf{f}) scaling.
    }
    \label{fig_main: sus mot}
\end{figure}

We next demonstrate that a short current pulse can trigger subsequent RF-driven DW motion even in the absence of a transverse symmetry-breaking field. The top part of Fig. \ref{fig_main: sus mot}a shows the applied current-pulse and RF-current sequence, while the lower panel defines their relative timing. Throughout this experiment, a longitudinal magnetic field \Hx{} is applied parallel to the DMI-stabilized Néel orientation of the DW. Unlike the transverse field \Hy{} used for self-propulsion (Fig. \ref{fig_main: self prop}), this longitudinal field does not rectify the RF-driven oscillation. Instead, it only modifies the magnitude of the current-induced DW displacement for both current polarities without breaking the left/right symmetry of the motion (see Methods and Supplementary Fig. \ref{fig-SI: v-Hxy}). We therefore do not expect an RF excitation alone to produce directed motion. Indeed, the top part of Fig. \ref{fig_main: sus mot}b shows that the mere application of RF-currents does not lead to an observable DW motion. We therefore investigated whether a current pulse perturbs the DW in a way that enables subsequent RF-driven motion. The remaining snapshots of Fig. \ref{fig_main: sus mot}b compare the DW displacement induced by repeated current pulses with that obtained from repeated pulse-RF sequences. When the RF burst precedes the current pulse, the pulse-induced DW displacement remains unchanged. By contrast, when the RF burst follows the pulse (orange frame), the DW moves significantly further. These observations show that the relative timing of current pulse and RF-current is crucial and that the initial pulse may act as a trigger that allows for subsequent RF-driven DW motion. The dependence on the RF phase at the time of the triggering current pulse was not investigated in this study. In the following, we identify the pulse conditions required to trigger the RF-driven motion.

The energy of the current pulse, defined by current density \jp{} and pulse width \pulseT{}, determines whether RF-driven motion is triggered or not. Fig. \ref{fig_main: sus mot}c displays the total DW displacement \dtotal{} $=$ \Xdisp{} $+$ \dRF{} versus current density, where \Xdisp{} denotes the displacement that occurs only due to the current pulse. It shows that without RF-currents, the displacement is symmetric in magnitude $($\dtotal{}$($\jp{}$) \approx |$\dtotal{}$(-$\jp{}$)|)$ and exhibits its typical saturation at high current densities. Upon application of an RF-current, \dtotal{} sharply increases once a threshold of \jp{} $\approx 108$ \MAcm{} is exceeded and this increase is larger for a longer RF-burst. It is noteworthy that the RF-driven DW motion still follows the direction of the pulse-induced DW motion. In addition to the threshold of the pulse current density, a sufficient pulse width is also required to achieve the RF-enhancement of DW motion. This is displayed in the inset of Fig. \ref{fig_main: sus mot}c, where the RF-enhancement becomes apparent for \mbox{\pulseT{} $\geq 3$ ns} and saturates for \pulseT{} $\geq 5$ ns. These results suggest that the current pulse needs to exceed threshold values in both current density and pulse duration to perturb the DW such that the RF-current induces sustained DW motion in the same direction as the current pulse. We therefore refer to this pulse-triggered, RF-driven DW motion as \textit{sustained} DW motion. 

In addition to the pulse requirements, the sustained DW motion requires sufficient RF power. Fig. \ref{fig_main: sus mot}d displays the DW displacement per pulse/RF combination and demonstrates that powers of \PowOutNew{} $\geq -4.5$ dBm are required for \dtotal{} to deviate from the baseline. Note that the RF-driven enhancement of \dtotal{} even occurs in zero-field, but it scales to much larger values for a finite field (Supplement Fig. \ref{fig-SI: sus-mot Hx frequency}a). The displacement scales linearly with the RF duration \tRF{} as shown in the inset. This implies that during the RF-burst the DW moves in a steady, continuous fashion. We would like to point out that the power threshold and the linear \tRF{} dependence are remarkably similar to self-propulsion, suggesting that both effects arise from a common rectification mechanism. 

Next, to understand the key relevance of the relative pulse-to-RF timing \DeltaT{} for the RF-driven motion, we gradually shifted the current pulse and the RF current with respect to each other. Fig. \ref{fig_main: sus mot}e confirms that the current pulse acts as the trigger for sustained motion, i.e. the enhancement of the DW displacement is maximum when the pulse reaches the device before the RF-burst and zero when it reaches afterwards. For intermediate \DeltaT{} between $-1$ \MuSec{} and $0$ \MuSec{}, the nanoseconds-long pulse arrives during the much longer RF-burst. Here, we see a linear increase of \dtotal{}, clearly showing that only the RF-duration after the arrival of the current pulse contributes to net motion of the DW. In the logarithmically-scaled Fig. \ref{fig_main: sus mot}f, we investigated the decay time of the initial current trigger, i.e. after what characteristic waiting time \DeltaT{} the sustained motion ceases. The data shows that the RF-driven motion is stable with almost no change up to \DeltaT{} $= 1$ \MuSec{}. The total displacement then drops steeply around \mbox{\DeltaT{} $\approx 10$ \MuSec{}} and no RF-effect is visible for \DeltaT{} $\geq 10$ ms. Such a long stability of the sustained motion does not match the characteristic timescales of magnetization dynamics which amount to nanoseconds and below. It is thereby clearly not related to DW inertia \cite{2010Thomas, 2016Torrejon}. However, the stability timescale of the pulse-induced perturbation of the DW agrees with the much slower pinning-dominated creep motion of magnetic textures \cite{2000Chauve, 2007YamanouchiCreep, 2010KandaCreep}. We will next address the nature of the pulse-induced perturbation of the DW and its role in the onset of sustained DW motion.

\subsection*{Microscopic Mechanism of Sustained Domain Wall Motion}
Specifically, we examine how the triggering current pulse modifies the magnetic configuration of the DW and how this, in turn, breaks the DW's symmetric response to a periodic RF current. Such symmetry breaking could, for instance, arise from current-induced DW tilting \cite{2018GargYshape} or curvature, which modify the equilibrium orientation of the magnetization within the wall.

\begin{figure}[]
    \centering
    \includegraphics[width=\linewidth]{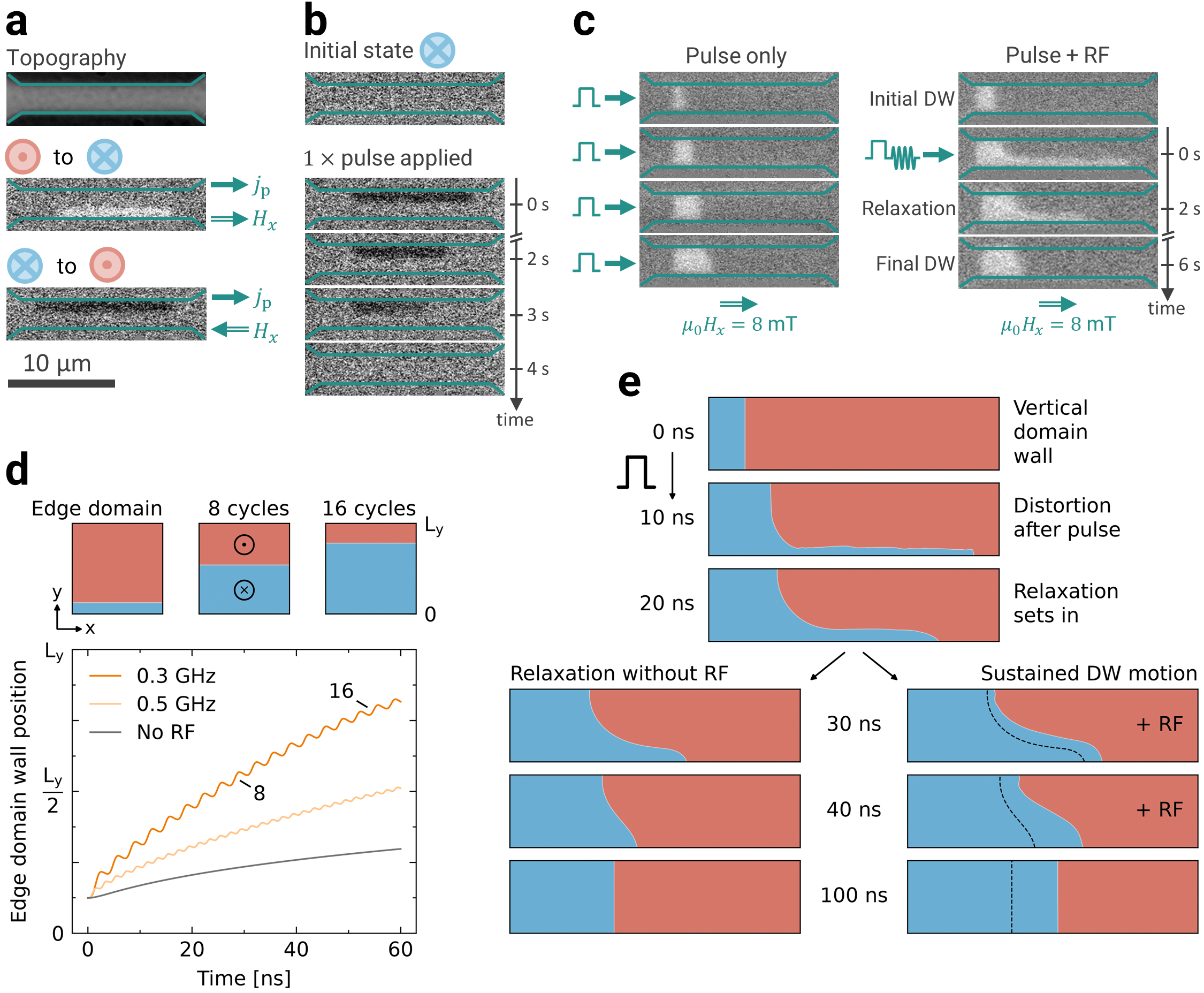}    
    \caption{
    \textbf{Pulse-induced edge-domain nucleation and RF-driven expansion during sustained DW motion.}
    (\textbf{a}) Pulse-induced nucleation of a metastable edge domain using \jp{} $=214$ \MAcm{} and \pulseT{} $=10$ ns in a field of \mbox{\muHx{} $=\pm16$ mT}. Green contours outline the racetrack and white (black) contrast indicates down (up) magnetization.
    (\textbf{b}) Gradual relaxation of the metastable edge domain after the end of the current pulse. The field still remains applied.
    (\textbf{c}) Impact of current pulses or a pulse-RF sequence on a DW. Current pulses alone translate the DW while preserving an approximately straight profile (left). When the pulse is followed by an RF burst \mbox{($f=100$ MHz,} \PowOutNew{} $=-2.1$ dBm, \tRF{} $=600$ ns) a switched edge domain becomes visible at the lower edge, merges with the DW and strongly distorts its shape and finally relaxes (right). 
    (\textbf{d}) Micromagnetic simulations of RF-driven expansion of an edge domain in a quadratic racetrack strip ($\mathrm{L_x} = \mathrm{L_y} = 3$ \MuMeter) with periodic boundary conditions along $x$ under a longitudinal field \mbox{\muHx{} $= 20$ mT}. The corresponding edge DW position shows oscillatory expansion under RF excitation \mbox{(\jRF{} $=20$ \MAcm{})} and slow DW drift without RF.
    (\textbf{e}) Simulations of pulse-induced edge switching and sustained DW motion in a large simulation box \mbox{($\mathrm{L_x} \times \mathrm{L_y} = 12 \times 3$ \MuMeter$^2$).} Without RF, the edge domain gradually relaxes (left), whereas subsequent RF excitation ($f=0.3$ GHz, \mbox{\jRF{} $=20$ \MAcm{}}) expands the edge domain and enhances the DW displacement (right).
    }
    \label{fig_main: sus mot explain}
\end{figure}

We first investigate the effect of a current pulse on a uniformly magnetized racetrack without a pre-existing DW. At comparatively high pulse amplitudes and finite longitudinal field \Hx{}, a reversed domain nucleates preferentially at one racetrack edge (Fig. \ref{fig_main: sus mot explain}a). The selected edge depends on the initial magnetization, the sign of \Hx{}, the DMI and the Oersted field, consistent with the established asymmetry of SOT-induced nucleation \cite{2015Martinez, 2018CaoSOTedge, 2017Baumgartner}. The pulse does not reverse the magnetization throughout the complete racetrack, and the resulting edge domain is metastable: it persists up to several seconds before collapsing back into the uniformly magnetized state \mbox{(Fig. \ref{fig_main: sus mot explain}b)}. This slow relaxation is consistent with pinning-dominated DW motion  \cite{2000Chauve, 2007YamanouchiCreep, 2010KandaCreep} driven by the DW surface tension that tries to minimize the DW area \cite{2018ZhangLaplace}.

Such metastable edge domains are observed only in devices fabricated by a lift-off method (see Methods), which produces gradually tapered edges that are easier to switch, presumably because of a locally reduced magnetic anisotropy that makes the edges magnetically softer (Supplementary Fig. \ref{fig-SI: AFM switching}). Etched devices instead show either no reversal or complete switching, without a comparable metastable intermediate state. To examine how magnetically softer edges affect the nucleation of edge domains, we performed micromagnetic simulations using MuMax3 \cite{2014Mumax}. The simulations reproduce the experimentally observed switching asymmetry between the top and bottom edges (Supplementary Fig. \ref{fig-SI: Mumax switching asymmetry}). These observations indicate that the triggering pulse can induce a metastable, spatially localized magnetic domain at the racetrack edge.

We next examine how this edge domain nucleation interacts with a DW. Under the lower-drive conditions typically used for DW motion, a current pulse alone leaves the optically resolved wall approximately straight (Fig. \ref{fig_main: sus mot explain}c, left). By contrast, the combined pulse–RF sequence reveals a visible switched edge region that merges with the wall, resulting in a strongly curved and elongated DW profile (Fig. \ref{fig_main: sus mot explain}c, right). After the excitation, the edge region gradually collapses and the wall recovers its approximately straight configuration due to the DW's surface tension. The absence of an optically resolved edge domain after the pulse alone does not exclude the formation of a smaller or shorter-lived reversed nucleus below the spatial and temporal resolution of the Kerr microscope. The experimental observations therefore suggest that the RF current expands and/or stabilizes a pulse-generated edge nucleus, allowing it to merge with the existing DW.

To test this interpretation, we perform two sets of micromagnetic simulations. First, we study the impact of an RF current on an edge domain. The simulation snapshots in Fig. \ref{fig_main: sus mot explain}d show that the initial edge domain expands from the bottom towards the top edge when exposed to an RF current. The corresponding position traces for $f=500$ MHz and $f=300$ MHz reveals that this expansion occurs through oscillatory DW motion and is faster at the lower frequency. This behavior is consistent with the low-frequency response of the DW established in Fig. \ref{fig_main: Intro} and resembles the self-propulsion discussed in Fig. \ref{fig_main: self prop}. Without RF excitation, the simulated edge wall slowly drifts towards the racetrack center owing to the symmetric simulation box, the absence of pinning sites and the periodic boundary conditions (see Methods), in contrast to the experimentally observed edge domains, which eventually collapse at the device edge. These simulations show that RF spin torques are capable of driving the expansion of an edge domain towards the opposite edge. This further suggests that pulse-generated edge domains, even when too small to be resolved by static Kerr microscopy, can be expanded by RF excitation to become experimentally observable.

We next investigate how such an expanding domain interacts with an initially present DW. Fig. \ref{fig_main: sus mot explain}e shows how a current pulse displaces an initially straight DW and simultaneously nucleates an edge domain at the lower edge. Without subsequent RF excitation, this region slowly collapses and thereby moderately drags the DW forward. When the pulse is followed by an RF excitation, the edge domain expands towards the upper edge. This RF-driven expansion progressively deforms and advances the wall beyond the displacement produced by the pulse alone (dashed line). Animations of both datasets are provided in the Supplementary Materials (see VS1 and VS2).
Micromagnetic simulations further show that sustained DW motion is most pronounced at low RF frequencies (Supplementary Fig. \ref{fig-SI: Mumax sus-mot freq-dep}), consistent with the experimental observations (Supplementary Fig. \ref{fig-SI: sus-mot Hx frequency}b). Moreover, a systematic parameter sweep of RF-driven edge-domain expansion (Supplementary Fig. \ref{fig-SI: Mumax sus-mot mechanism}) demonstrates how the RF current amplitude, RF frequency, and \Hx{} jointly determine the expansion efficiency. We note that these experimental and numerical results indicate that the current pulse prepares a metastable edge domain, while the subsequent RF excitation drives its expansion, sustaining the motion of the connected DW.

\subsection*{Evolution from Sustained Domain-Wall Motion to RF-Assisted Switching}
\begin{figure}[h!]
    \centering
    \includegraphics[width=\linewidth]{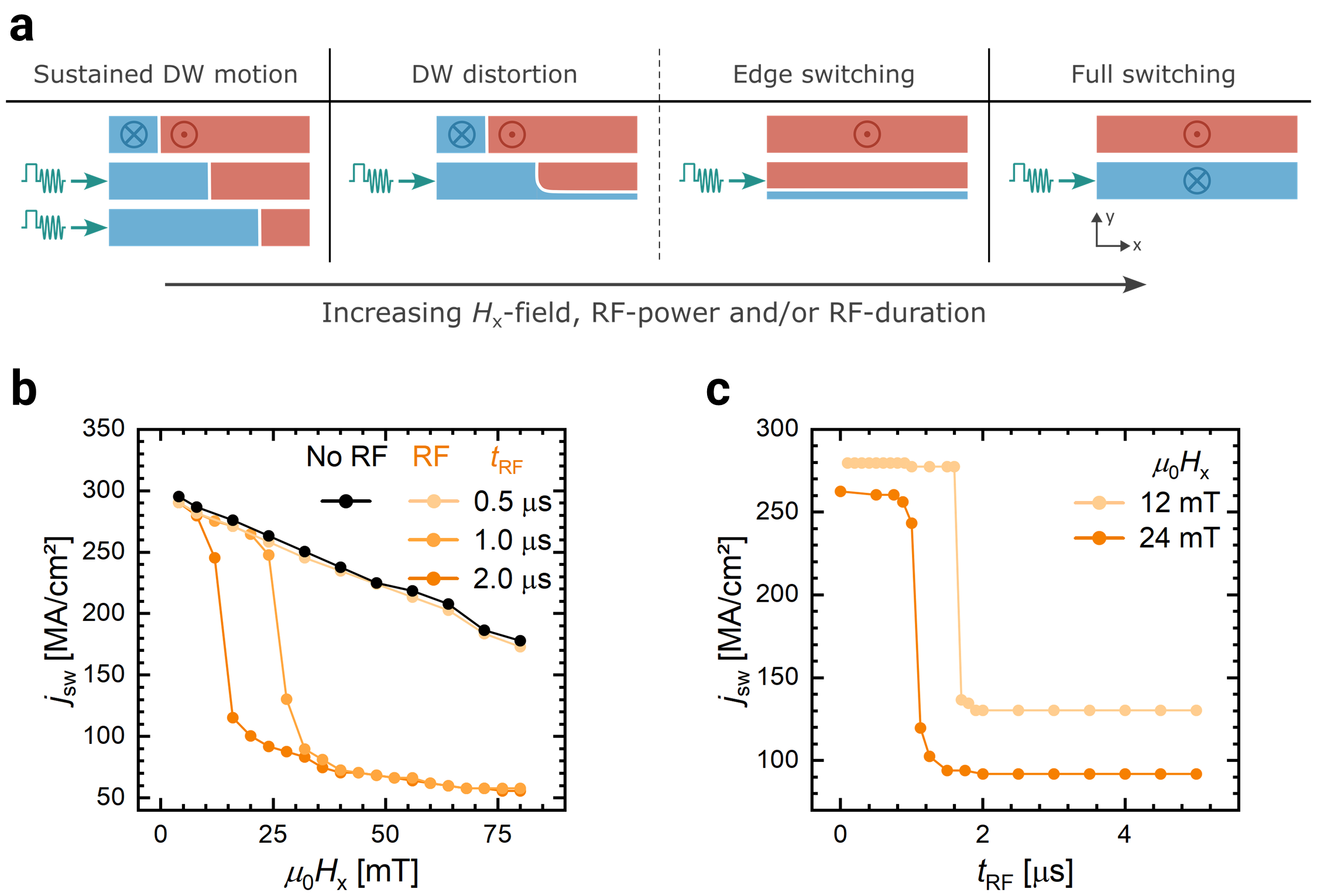}    
    \caption{\textbf{From sustained domain wall motion to magnetic switching.}    
    (\textbf{a}) Schematic evolution from sustained DW motion (left) through an intermediate regime characterized by DW distortion and edge switching (center) to RF-assisted switching (right). Increasing the \Hx{} field and the strength of the RF excitation (by increasing the RF power or duration) progressively transforms sustained propagation into complete magnetic reversal. 
    \mbox{\Hx{}-dependence (\textbf{b})} and \mbox{\tRF{}-dependence \textbf{(c)}} of the switching current density \jsw{}:
    An initial current pulse with \pulseT{} $=5$ ns and a subsequent RF-burst with $f=100$ MHz, \PowOutNew{} $= -2.8$ dBm, $\Delta t \approx 10$ ns are applied to reverse the magnetization from up to down. The 'No RF' data in \textbf{(b)} is shifted up by 5 \MAcm{} to enhance visibility of the underlying data trace.    
    Together, these observations show that increasing the RF-driven expansion of pulse-generated edge domains continuously transforms sustained DW motion into complete magnetic switching.
    }
    \label{fig_main: switching}
\end{figure}

The experiments and simulations in Fig. \ref{fig_main: sus mot explain} indicate that sustained DW motion originates from RF-driven expansion of a pulse-generated edge domain. In this framework, we expect that a further RF-driven expansion of the edge-domain should ultimately result in complete magnetization reversal. Fig. \ref{fig_main: switching}a summarizes this proposed evolution from sustained DW motion to RF-assisted magnetic switching. For weak RF excitation and low longitudinal field \Hx{}, the RF-driven expansion remains limited and manifests itself as sustained DW motion with an approximately straight wall profile (left). Increasing \Hx{} or strengthening the RF excitation enlarges the edge domain until it becomes visible (centre), leading to the distorted DW configurations as previously discussed in Fig. \ref{fig_main: sus mot explain}c. Under even stronger excitation, the edge domain expands across the entire racetrack width, resulting in complete RF-assisted switching (right). The corresponding experimental microscopy images are summarized in Supplementary Fig. \ref{fig-SI: evolution sus-mot switching}. These results establish a continuous evolution from sustained DW motion to RF-assisted switching, identifying both phenomena as two limits of the same RF-driven edge-domain expansion process. We now quantify this transition by examining the switching current required for complete RF-assisted reversal.

The field dependence of the required switching current density (\jsw{}) of the current pulse is dramatically modified by the presence of an additional RF current. Here, full switching refers to magnetization reversal across the entire racetrack width. Without RF excitation, \jsw{} decreases linearly with increasing \Hx{} which is consistent with SOT-switching \cite{2013Lee}. With an additional RF-current, a sudden drop of \jsw{}, followed by a gradual flatting, is observed. For instance, at $40$ mT, an additional RF-current with \tRF{} $=1.0$ \MuSec{} reduces the required \jsw{} from \mbox{$233$ \MAcm{}} (no RF) to $73$ \MAcm{} (RF). The RF-induced reduction of \jsw{} is highly non-linear in \Hx{}. Increasing \tRF{} lowers the field at which this sharp drop in \jsw{} occurs, but when \tRF{} is too short, no RF-induced reduction in \jsw{} is observed. Together with the mechanism established in Fig. \ref{fig_main: sus mot explain}, these observations are consistent with the current pulse first generating a sufficiently large and stable edge domain, which the subsequent RF excitation expands across the racetrack. The observed \Hx{} threshold is therefore naturally explained by the fact that increasing \Hx{} promotes both the formation of the edge domain and its subsequent RF-driven expansion. Note that an abrupt reduction of \jsw{} is also observed in the \tRF{} dependence in Fig. \ref{fig_main: switching}c. For \mbox{\muHx{} $=12$ mT}, the switching threshold drops by $141$ \MAcm{} when the RF-duration in increased from $1.6$ \MuSec{} to $1.7$ \MuSec{}. Before and after this drop, \jsw{} remains mostly constant. Increasing \Hx{} reduces the required RF duration to induce RF-assisted switching. Based on the simulations in Fig. \ref{fig_main: sus mot explain}d, the RF-driven edge domain expansion requires a finite time. Likewise, the reduction of the \tRF{} threshold with increasing \Hx{} is consistent with a faster RF-driven edge domain expansion in larger \Hx{} fields (Supplementary Fig. \ref{fig-SI: Mumax sus-mot mechanism}d). Additional parameter dependences of \jsw{} in Supplementary Fig. \ref{fig-SI: RF switching params} show trends consistent with the mechanism discussed here.
\newpage

\newpage

\subsection*{Conclusion}
Our results establish that a DW oscillation induced by RF currents can be converted into directed DW transport through symmetry breaking. Such symmetry breaking may be introduced either by external magnetic fields or through pulse-induced perturbations of the magnetic configuration, giving rise to self-propulsion, sustained DW motion and, ultimately, RF-assisted magnetic switching. While self-propulsion arises from the direct rectification of the RF-driven DW oscillation, sustained motion originates from the RF-driven expansion of a pulse-generated edge domain. This expansion therefore represents an alternative manifestation of RF-driven rectification and naturally connects sustained DW motion with RF-assisted magnetic switching. Together, these results establish RF-driven rectification as a unifying framework for understanding directed DW motion and RF-assisted switching. More generally, the periodic excitation of non-uniform magnetic textures and the resulting rectification of their motion represent a fundamental concept that is not restricted to spin-orbit torques or chiral Néel DWs. A similar behavior is expected for other driving mechanisms and across a broad range of magnetic textures and materials, providing new opportunities for the high-frequency control of magnetic textures and future spintronic devices.

\newpage

\clearpage 
\bibliography{bibliography}
\bibliographystyle{sciencemag}

\subsection*{Materials and Methods}

\subsubsection*{Device fabrication and thin film preparation}
The measured devices were fabricated on highly-insulating floating-zone Silicon substrates with 001-cut using a two step photo-lithography process: First, using a double-layer resist out of ARBR5480/ARN4340 (negative resist from 'Allresist GmbH'), a resist mask with pronounced undercut was formed. Then the magnetic thin films were deposited using magnetron sputtering to form the magnetic racetrack. After the deposition of the magnetic structures and lift-off of the resist, a conventional lift-off technique was used to deposit electrical contact pads by ion-beam deposition (5 nm Ti / 60 nm Au). The straight part of the magnetic racetrack has a length of $20$ \MuMeter{}, a width of $3$ \MuMeter{} and a gap of $5$ \MuMeter{} to the ground planes. The ground planes have a width of $w_\mathrm{G} = 100$ \MuMeter{} in the center region. These dimensions laterally increase towards the outer regions where the wirebonds are placed for electrical contacting.

The magnetic thin films were grown by a home-built d.c. magnetron sputtering at an Argon pressure of \mbox{$3$ mTorr}. The base pressure of the home-built sputtering system was below \mbox{10$^{-9}$ Torr}. The film structure was as follows: 20 TaN / 50 Pt / 0.9 Ta / 3 Co / 7 Ni / 3 Co / 30 TaN (the numbers indicate the thickness in $\mathring{\textrm{A}}$ngström). Note that the Tantalum insertion layer between Pt and Co tunes, i.e. reduces, the magnetic anisotropy and the DMI of the magnetic film - which strongly depends on the Pt/Co interface.

The static and dynamic magnetic properties of the magnetic films were measured using vibrating-sample magnetometry (VSM) and FMR, respectively (see Supplementary Figs. \ref{fig-SI: VSM}, \ref{fig-SI: FMR}). Note that these measurements were performed on unpatterned $5 \times 5$ mm² samples at room temperature. A home-built FMR setup was used to measure the resonant frequency, field linewidth and damping constants of the magnetic thin film. In order to perform the measurement, the sample was placed on top of a co-planar waveguide (Rosenberger GmbH). Passing microwaves with GHz frequencies through the co-planar waveguide generates a high-frequency magnetic field in the plane of the magnet. This excitation field is perpendicular to a static field of an electromagnet that is oriented along out-of-plane along the easy axis of the magnet. The amplitude of the static field is weakly modulated and hence modulates the microwave absorption of the sample at the FMR. After rectification of the microwave signal, it is detected with a lock-in amplifier.

\subsubsection*{Super-Nyquist sampling magneto-optical Kerr microscopy}
To investigate the dynamic magnetic response of the DWs and the adjacent uniform domains, we utilized a variant of time-resolved magneto-optical Kerr (TR-MOKE) microscopy that we refer to as Super-Nyquist sampling magneto-optical Kerr microscopy (SNS-MOKE) \cite{2021Dreyer}. In our approach, an electrical RF current at MHz or GHz frequencies passes through the micro-structured magnetic racetrack, thereby exerting an RF spin torque on the magnetization. As an optical probe, we employed a short-pulsed laser system operating at 515 nm with a repetition frequency of $f_\mathrm{rep} = 80$ MHz. By means of an high-numerical aperture microscope objective the laser beam is focused to a diffraction limited spot size of $\approx$ 300 nm. Thus we obtain a signal which corresponds to the dynamic change of the light's polarization state under reflection via the magneto-optical Kerr effect within the diffraction limited area of the laser spot.  

In contrast to conventional TR-MOKE experiments, our approach does not require additional external modulation techniques. To achieve this, we stabilize the laser's repetition rate, the frequency generator as well as the Lock-in amplifier to an external master clock. This allows for the direct demodulation at alias frequency components $|n \cdot f_\mathrm{rep} - f_\mathrm{RF}|$, where $n$ is an integer number representing the comb lines of the frequency comb of the ultra-short laser pulses. These alias frequencies originate from an under-sampling of the magnetization dynamics at $f_\mathrm{RF}$ with the 80 MHz sampling frequency of the laser pulses. However, since all frequency components are stabilized to the external clock, the exact position of the alias frequency is predetermined and can be used for demodulation. This case of under-sampling allows to reconstruct the amplitude and phase information at arbitrary frequencies. 

In this work we distinguish between two different MOKE geometries:
(1) For out-of-plane magnetized samples, as demonstrated in this work, the dynamic P-MOKE component only becomes visible in the region of an oscillating DW. Inside the probed beam area, the average $z$-component of the magnetization is periodically changing, leading to a finite P-MOKE signal. In contrast, the P-MOKE is not sensitive to FMR since all periodically varying magnetization components lie in-plane in this case.
(2) In order to detect the FMR and image the precessional motion of an out-of-plane magnetized film, one needs to use L-MOKE to access the dynamic longitudinal component of the magnetization which points along $x$-direction (i.e. in-plane and along the racetrack axis). For this, we block half of the incoming beam (Supplementary \Fig{} \ref{fig-SI: Illumination_path}) and cut half of the wave-vector spectrum provided by the microscope objective. As a consequence of this, the spatial resolution is slightly reduced. However, the longitudinal components of the Kerr rotation will not average to zero anymore and can be obtained by means of lock-in detection. In such a geometry, the L-MOKE is sensitive to the in-plane dynamics in the domains, where the magnetization points out-of-plane and the precession has $x$- and $y$-components, but not to the dynamics within the DW.

\subsubsection*{Domain wall motion measurement}
In order to investigate DW motion, static Kerr microscopy in P-MOKE geometry was used. In particular, differential mode imaging, i.e., by initially subtracting a background image, yields a significantly enhanced contrast for moving DWs. A change of the $z$-component of the magnetization increases or decreases the detected Kerr intensity relative to the background image. An up-to-down change of the local magnetization therefore appears white, while a down-to-up change appears as black. 

For the measurements of DW motion, we first isolated a single DW in the racetrack. A static magnetic in-plane field in transverse or longitudinal orientation, relative to the racetrack axis, can be applied. The in-plane magnetic field was carefully aligned with the sample plane using a custom-built magnet holder. The alignment was verified by applying nominally in-plane fields of up to $160$ mT, for which no DW displacement was observed. Based on an independently measured out-of-plane DW depinning field of $\mu_0 H_\mathrm{dep} = 1.15$ mT (Supplementary \Fig{} \ref{fig-SI: OOP depinning field}), we estimate an upper bound on the angular misalignment of $\theta < 0.41^\circ$.

In addition to a magnetic field, the DW can be manipulated by a pulsed current, an RF current or a combination of both. A nanosecond pulse generator (PSPL 10300B or AVTECH AVR-E3) is used to apply a series of current pulses with current density \jp{} and width \pulseT{}. An RF signal generator (Keysight N5183B MXG) generates a continuous signal which is chopped into RF-bursts of length \tRF{} using a microwave mixer (Marki Microwave M1-0012LQP) and a burst pulse generator. The bursts are then amplified using an RF amplifer (MiniCircuits ZVA-0.5W303G+). A resistive power combiner adds the pulse and RF signals and feeds them into the racetrack. The transmitted signal is monitored by an oscilloscope (Tektronix MSO64). The relative timing of current pulse and RF signal is precisely set by a digital delay and pulse generator (DG535 Stanford Research Systems) that triggers the pulse and burst generators. The effect of the RF phase at the time of the current pulse was not investigated in this study. A circuitry schematic is shown in Supplementary Fig. \ref{fig-SI: circuitry}.

When DW motion is induced by current pulses or RF currents, the total DW displacement \dtotal{}, relative to the background image, becomes visible in the differential image. Multiple pulse-RF combinations are applied to increase the total displacement and improve the measurement accuracy by averaging the displacement per pulse-RF combination. Furthermore, this measurements procedure is repeated several times to obtain the statistical mean and standard deviation of the DW displacement.

\subsubsection*{Current-induced DW motion and typical field dependences}
An electrical current flowing through the magnetic racetrack induces an SOT acting on the local magnetization. This SOT is the major source for current-induced DW motion \cite{2013Emori, 2013Ryu, 2011Miron} in our magnetic system and moves the DW opposite to the electron flow. In zero magnetic field, this DW displacement \Xdisp{} is symmetric upon reversal of the current polarity, i.e. \Xdisp{}$(+$\jp{}$)=|$\Xdisp{}$(-$\jp{}$)|$. In the presence of a static in-plane magnetic field, however, the current-induced DW motion is significantly modified.

A static transverse \Hy{} field rotates the DW magnetization from a Néel toward a Bloch orientation and thereby makes the DW's response to current-induced SOT asymmetric. The DW displacement for positive and negative current polarity is now different, i.e. \Xdisp{}$(+$\jp{}$)\neq|$\Xdisp{}$(-$\jp{}$)|$ \cite{2013Ryu} (Supplementary Fig. \ref{fig-SI: v-Hxy}a). For instance, a positive \Hy{} leaves \Xdisp{}$(+$\jp{}$)$ unchanged or enhances it, whereas $|$\Xdisp{}$(-$\jp{}$)|$ is strongly suppressed and eventually drops to zero at high \Hy{}. For a negative \Hy{}, the situation reverses such that \Xdisp{}$(+$\jp{}$)$ is suppressed while $|$\Xdisp{}$(-$\jp{}$)|$ is favored. Note that the impact of \Hy{} on current-induced DW motion is independent on the DW configuration (up-down or down-up DWs) \cite{2013Ryu}. Also note that no in-plane magnetic field alone can induce continuous motion of the DW.

In contrast, a static transverse \Hx{} field preserves the symmetry of DW motion but the absolute displacement is enhanced (reduced) if \Hx{} points along (opposite) the effective DMI-field \Hdmi{} of the chiral Néel DW. For the down-up DW in Fig. \ref{fig-SI: v-Hxy}b, \Hdmi{} points along $+x$ and the linear increase of $|$\Xdisp{}$|$ with increasing \Hx{} is apparent. For negative \Hx{}, i.e. when \Hx{} is opposing \Hdmi{}, $|$\Xdisp{}$|$ linearly decreases and eventually becomes zero for \mbox{\Hx{} $= -$\Hdmi{}}. The DW moves again for \Hx{} $<-$\Hdmi{} but the displacements reverse sign, i.e. the DW now moves along the electron flow instead of the current flow. Note that an up-down DW has the opposite \Hdmi{} field.

\subsubsection*{Collective coordinate model $q-\psi-\chi$}
To numerically calculate the DW dynamics in response to an RF current, we used the $q-\psi-\chi$ model that treats the DW as a one-dimensional object that is described by a position $q$, an in-plane tilting angle $\psi$ and a DW tilt $\chi$ \cite{2002Thiaville, 2012Thiaville, 2013Boulle}. The tilt coordinate is introduced to account for the significant DMI-induced tilting of a DW \cite{2012Ryu} in extended, micron-wide microwires during current-induced motion. Based on a Rayleigh-Lagrange approach, Boulle et al. \cite{2013Boulle} derived the $q-\psi-\chi$ model, i.e. the differential equations for $dq/dt$, $d\psi/dt$ and $d\chi/dt$. These equations are numerically solved using an implicit Runge-Kutta method to obtain the time trace of the DW position, angle and tilt. 

We simulated a down-up DW in a magnetic material with $M_\mathrm{S}=660 \cdot 10^3$ A/m, $K_\mathrm{u}=3.8 \cdot 10^5$ J/m$^3$, $D=-0.5\cdot 10^{-3}$ J/m$^2$, $\alpha = 0.045$, $\gamma/(2\pi)=28$ GHz/T. A racetrack width of $1$ \MuMeter{} and demagnetizing factors $N_\mathrm{x}=0.08$, $N_\mathrm{y}=0$ are considered. The spin Hall angle amounts to 0.1.

For the simulations on DW self-propulsion, a static \Hy{} is applied and an initial relaxation procedure is performed. Then, an additional RF-current with $f=100$ MHz and \jRF{} $=10$ \MAcm{} is applied. To obtain the DW drift velocity during self-propulsion, an adaptive solver was developed. It determines when a quasi-periodic state is reached, i.e. where the net DW drift per RF cycle converge to a constant value and where the $\psi$-variation becomes periodic with negligible cycle-to-cycle drift. Upon reaching this quasi-periodic state, the latest DW drift velocity is extracted. Note that the results on self-propulsion are qualitatively very similar when only a simplified $q-\psi$ model, i.e. without simulation of a DW tilt angle, is considered.

\subsubsection*{Micromagnetic simulations}
Micromagnetic simulations were performed using the \textit{mumax3} software \cite{2014Mumax}. The magnetic properties were either experimentally determined as $M_\mathrm{sat} = 660 \cdot 10^3$ A/m, $K_\mathrm{u1} = 3.8 \cdot 10^5$ J/m$^3$ or estimated as \mbox{$A_\mathrm{ex} = 1.5 \cdot 10^{-11}$ J/m}, $\alpha = 0.045$, $D_\mathrm{ind} = 0.5 \cdot 10^{-3}$ J/m$^2$. Pure damping-like SOT was added by adapting the existing Slonczewski spin transfer torque implementation \cite{2023JoosMumax} using a spin Hall angle of $0.1$. If not mentioned otherwise, all simulations were performed under a longitudinal field \muHx{} = 20 mT. For completeness, the current-induced Oersted-field contribution was also included by integrating the Biot-Savart law over the rectangular cross-section of the conductive layers of the film stack, i.e. over the cross-section of the Platinum and ferromagnetic layers. 

For the simulations on sustained DW motion in Fig. \ref{fig_main: sus mot explain}e and Supplementary Fig. \ref{fig-SI: Mumax sus-mot freq-dep}, we simulated a $L_\mathrm{x} \times L_\mathrm{y} = 12 \times 3$ \MuMeter{}$^2$ magnetic racetrack, discretized into a $4096 \times 1024$ grid. To mimic magnetically softer edges, the magnetic anisotropy $K_\mathrm{u1}$ was reduced linearly by 30\% within a 50 nm-wide region along the horizontal edges. A 2 ns current pulse with a current density of \jp{} $=105$ \MAcm{} displaced the initial vertical down-up DW and partially switched the bottom-right edge from up to down. After 20 ns, when the resulting distorted magnetization state started to relax, an RF current with $f=0.3$ GHz and \jRF{} $=20$ \MAcm{} was applied for $50$ ns. 

To study the RF-driven edge domain expansion in Fig. \ref{fig_main: sus mot explain}d, we used a square racetrack with $L_\mathrm{x} = L_\mathrm{y} = 3$ \MuMeter{}, the same cell size and and magnetically soft edges as before, and added periodic boundary conditions along the $x$-dimension. A horizontal down-up DW was initialized close to the bottom edge and an RF current with an amplitude of \jRF{} $=20$ \MAcm{} was applied for $60$ ns. For the study of asymmetric edge domain nucleation in Supplementary Fig. \ref{fig-SI: Mumax switching asymmetry}, we used the same square racetrack while varying the edge softening approaches, Oersted-field contribution and periodic boundary conditions to compare injection thresholds and dynamics. For the detailed parameter sweep of RF-driven edge domain expansion in Supplementary Fig. \ref{fig-SI: Mumax sus-mot mechanism}, we further simplified the geometry to a short, wide strip with $L_\mathrm{x} = 375$ nm and $L_\mathrm{y} = 6$ \MuMeter{}, also applying periodic boundary conditions along $x$. RF current was applied for $50$ ns and the mean domain expansion growth rate was evaluated between $10$ ns and $50$ ns to minimize contributions from the initial acceleration. No DW got closer than 1 \MuMeter{} to the top edge, so edge repulsion effects were minimal.

\newpage

\section*{Acknowledgments}
\paragraph*{Funding:}
Funding by the International Max Planck Research School for Science and Technology of Nano-Systems is gratefully acknowledged. 
Funding by Samsung Electronics Co., Ltd. is gratefully acknowledged. 
Funded by the European Union (ERC Advanced Grant SUPERMINT, project number 101054860). 
Financial support from the German Research Foundation (DFG) via Collaborative Research Center CRC/TRR 227 in project B10 and German Excellence Strategy – EXC3112/1 – 533767171 (Center for Chiral Electronics) is gratefully acknowledged.
Funded by the Deutsche Forschungsgemeinschaft (DFG, German Research Foundation) – project number 534336837.
\paragraph*{Author contributions:}
S.S.P.P. and G.W. directed the project. L.F., J.J., G.W., S.S.P.P. conceived the project and designed the experiment. J.J. prepared the films and L.F. characterized the films and fabricated the devices. R.D. performed the SNS-MOKE measurements. L.F. performed the static Kerr microscopy for the measurements of domain wall motion and switching. S.A. and L.F. conceived the numerical simulation. S.A. conducted and analysed all Mumax simulations and L.F. performed the collective coordinate model simulations. All authors discussed the results and prepared the manuscript.
\paragraph*{Competing interests:}
There are no competing interests to declare.

\section*{Supplementary materials}
Figs. S1 to S15 \\
Movies VS1 and VS2 

\newpage
\renewcommand{\thefigure}{S\arabic{figure}}
\renewcommand{\thetable}{S\arabic{table}}
\renewcommand{\theequation}{S\arabic{equation}}
\renewcommand{\thepage}{S\arabic{page}}
\setcounter{figure}{0}
\setcounter{table}{0}
\setcounter{equation}{0}
\setcounter{page}{1}

\begin{center}
\section*{Supplementary Materials for\\ \scititle}
Lukas Fischer$^{1}$, 
Rouven Dreyer$^{2,3}$, 
Simon Arnold$^1$,
Jae-Chun Jeon$^{1}$,\\
Georg Woltersdorf$^{1,2,3}$, 
Stuart S. P. Parkin$^{1\ast}$\\
\small$^{1}$Max Planck Institute of Microstructure Physics, Weinberg 2, 06120 Halle (Saale), Germany.\\
\small$^{2}$Department of Physics, Martin Luther University Halle-Wittenberg,  
\small Von-Danckelmann-Platz 3, 06120 Halle, Germany.\\
\small$^{3}$Cluster of Excellence - Center for Chiral Electronics, Halle-Berlin-Regensburg, Germany. \\
\small$^\ast$Corresponding author. E-mail: stuart.parkin@mpi-halle.mpg.de
\end{center}

\subsubsection*{This PDF file includes:}
Figures S1 to S15
\subsubsection*{Other Supplementary Materials for this manuscript:}
Movies VS1 to VS2\\

\newpage

\begin{figure}
        \centering
        \includegraphics[width=0.9\linewidth]{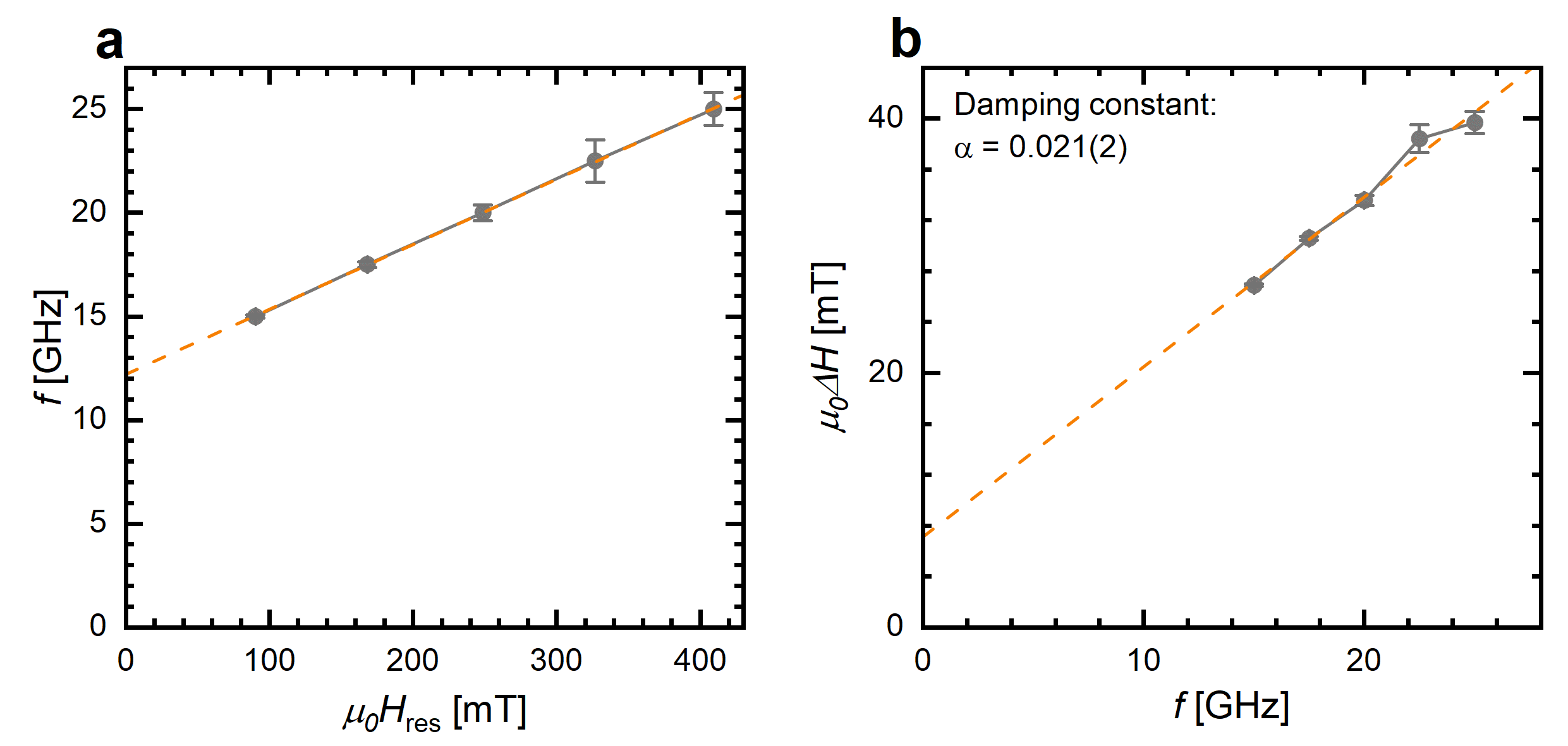}
        \caption{\textbf{Ferromagnetic Resonance.} Dependence of the FMR frequency $f$ on the static out-of-plane field. The gyromagnetic ratio $\gamma$ and the zero-field frequency $f_0$ were extracted from a linear fit $f_\mathrm{FMR} = f_\mathrm{0} + \gamma\cdot H_\mathrm{OOP}$. The linewidth analysis (inset) reveals the inhomogeneous broadening and the damping constant $\alpha$.}
        \label{fig-SI: FMR}
\end{figure}

\begin{figure}
    \centering
    \includegraphics[width=0.9\linewidth]{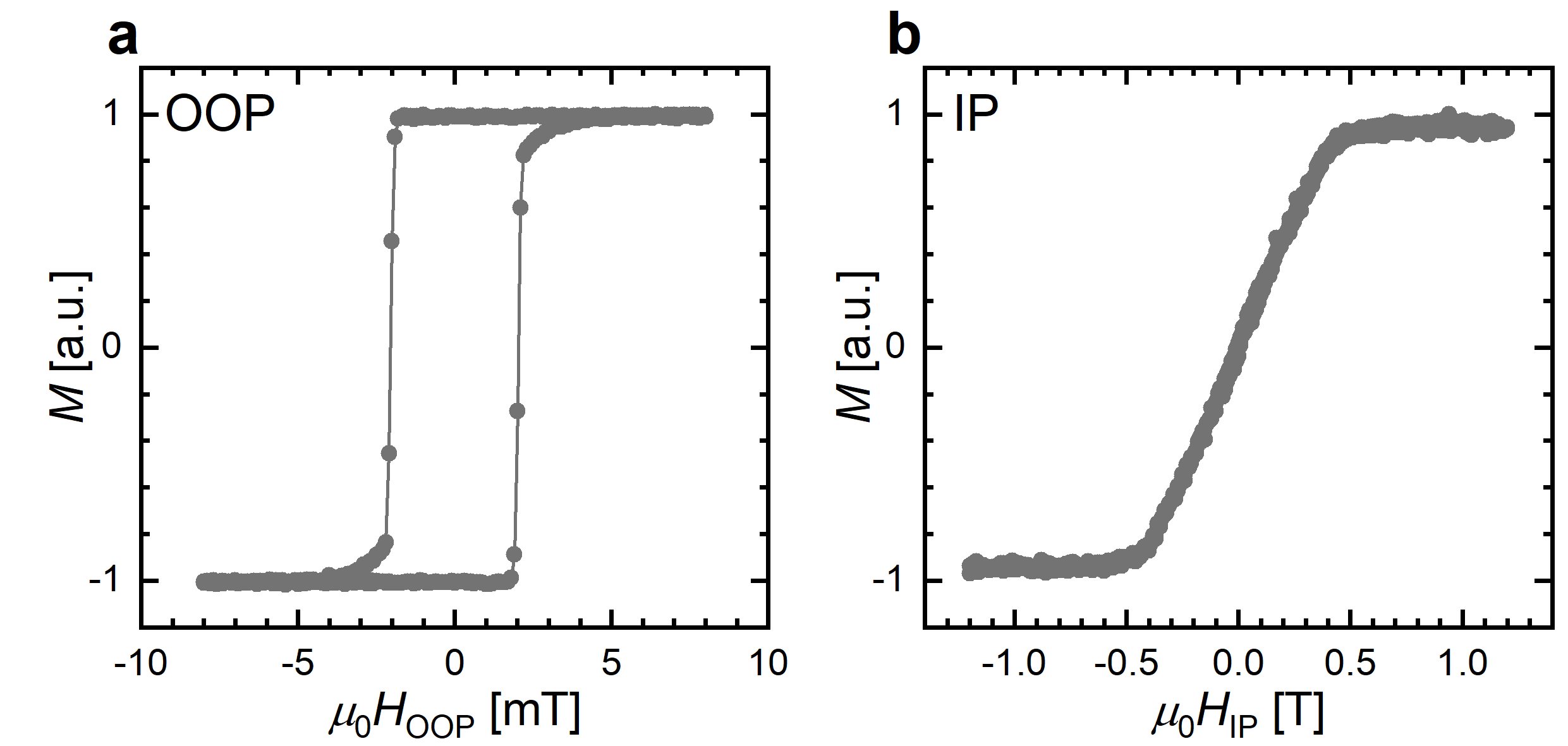}
    \caption{\textbf{Vibrating-sample magnetometry.} (\textbf{a}) Out-of-plane (OOP) magnetization $M_\mathrm{OOP}$ in an OOP magnetic field $H_\mathrm{OOP}$. (\textbf{b}) In-plane (IP) magnetization $M_\mathrm{IP}$ in an IP magnetic field $H_\mathrm{IP}$. The measurements were performed on a 8600 Series VSM by Lake Shore Cryotronics. The magnetization has been normalized.}
    \label{fig-SI: VSM}
\end{figure}

\begin{figure}
    \centering
    \includegraphics[width=\linewidth]{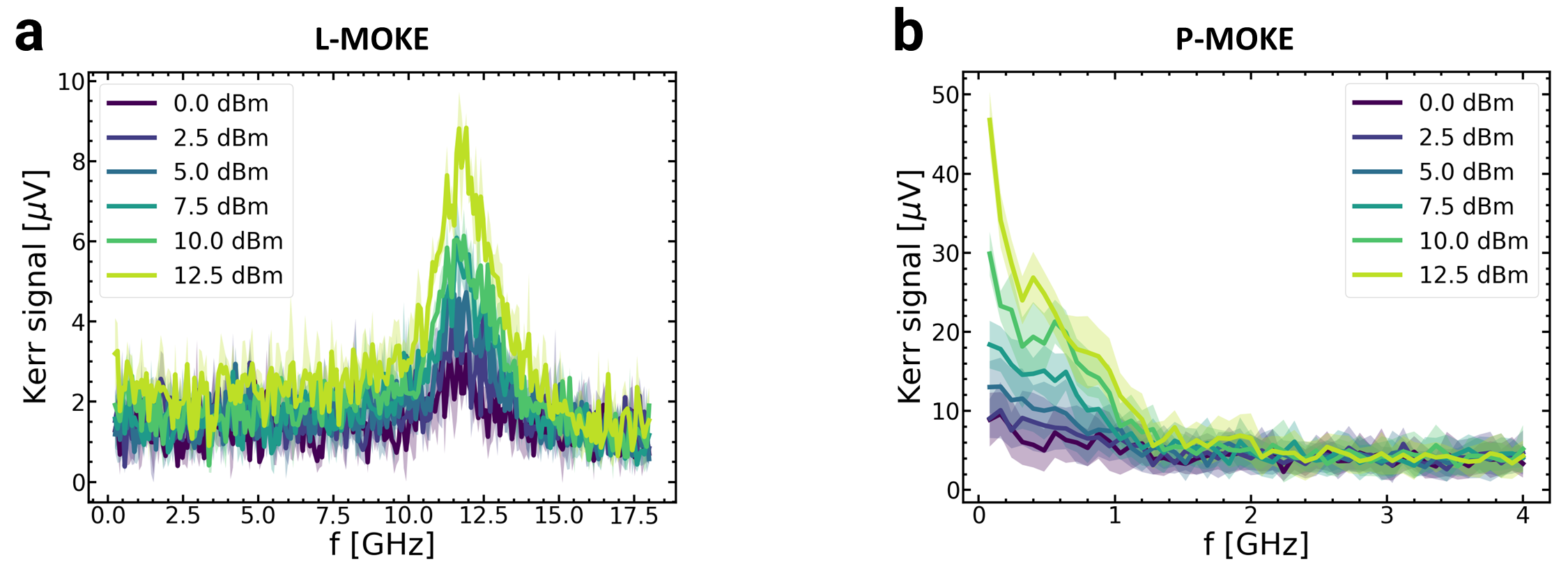}
    \caption{\textbf{RF power-dependences of SNS-MOKE measurements for FMR mode and DW oscillation.} 
    \textbf{(a)} Longitudinal MOKE measured in a uniform magnetic domain. 
    \textbf{(b)} Perpendicular MOKE measured at the DW position. 
    The specified RF powers are applied to the input of the device.
    }
    \label{fig-SI: SNS power dep}
\end{figure}

\begin{figure}
    \centering
    \includegraphics[width=0.7\linewidth]{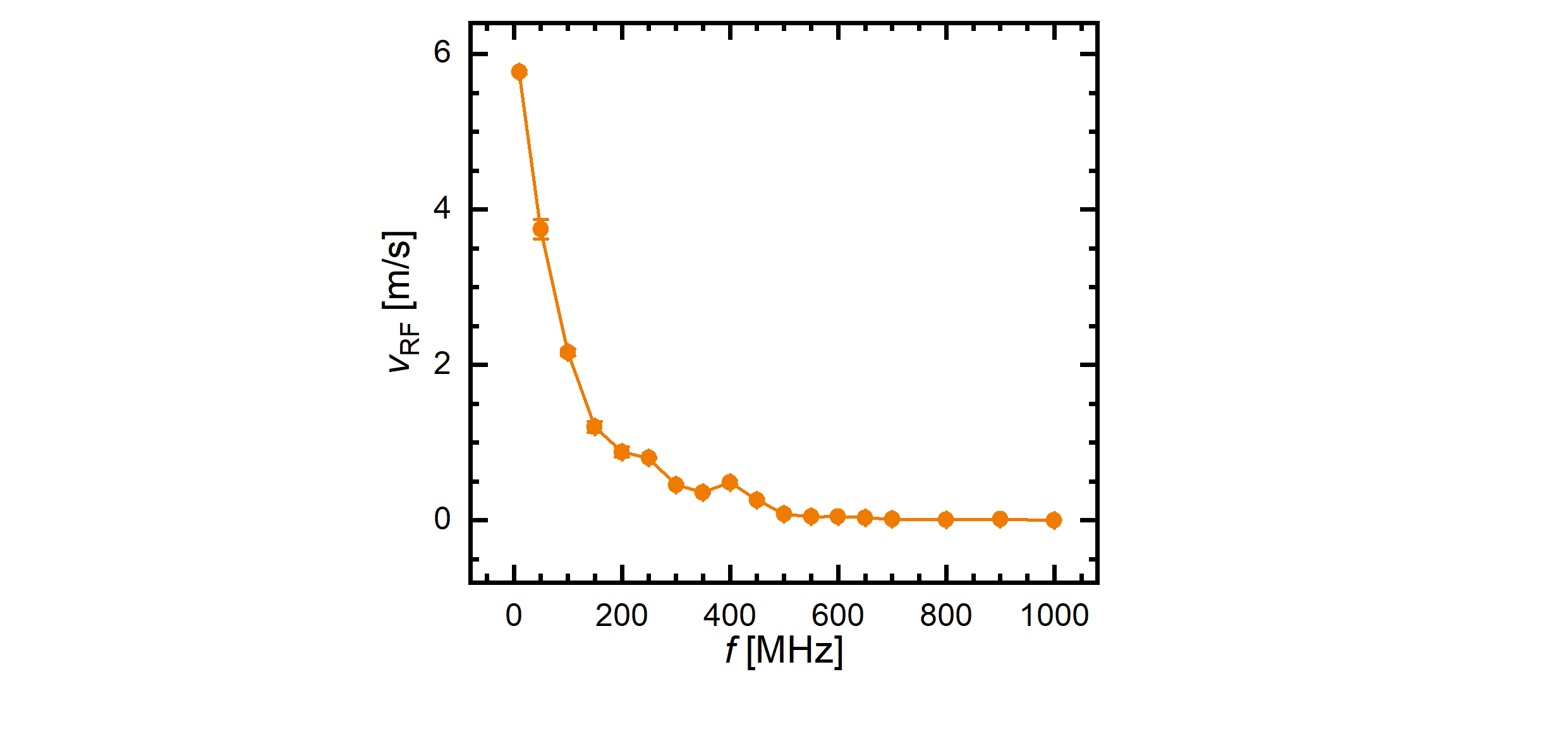}
    \caption{\textbf{RF frequency dependence of self-propulsion.} 
    An RF current with a constant output power of \mbox{\PowOutNew{} $=-3$ dBm} is applied in a transverse field \muHy{} $=40$ mT.
    }
    \label{fig-SI: self-prop frequency}
\end{figure}

\begin{figure}
    \centering
    \includegraphics[width=\linewidth]{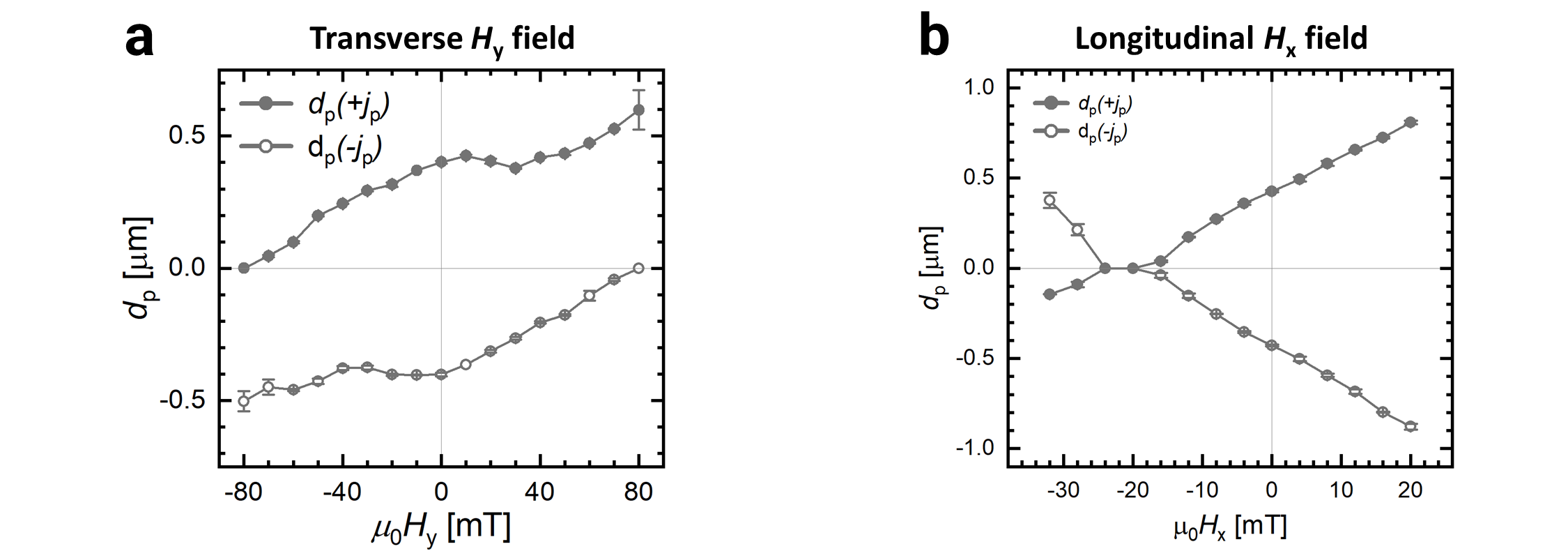}
    \caption{\textbf{Typical field-dependencies \Xdisp{}(\Hy{}, \Hx{}) of the current-induced motion of a down-up DW.} (\textbf{a}) Transverse field dependence \Xdisp{}(\Hy{}) using \mbox{\jp{} $= 107.5$ \MAcm{}}.
    (\textbf{b}) Longitudinal field dependence \Xdisp{}(\Hx{}) using \mbox{\jp{} $= 136$ \MAcm{}}.
    Current pulses with \mbox{\pulseT{} $= 5$ ns} are applied.}
    \label{fig-SI: v-Hxy}
\end{figure}

\begin{figure}
    \centering
    \includegraphics[width=\linewidth]{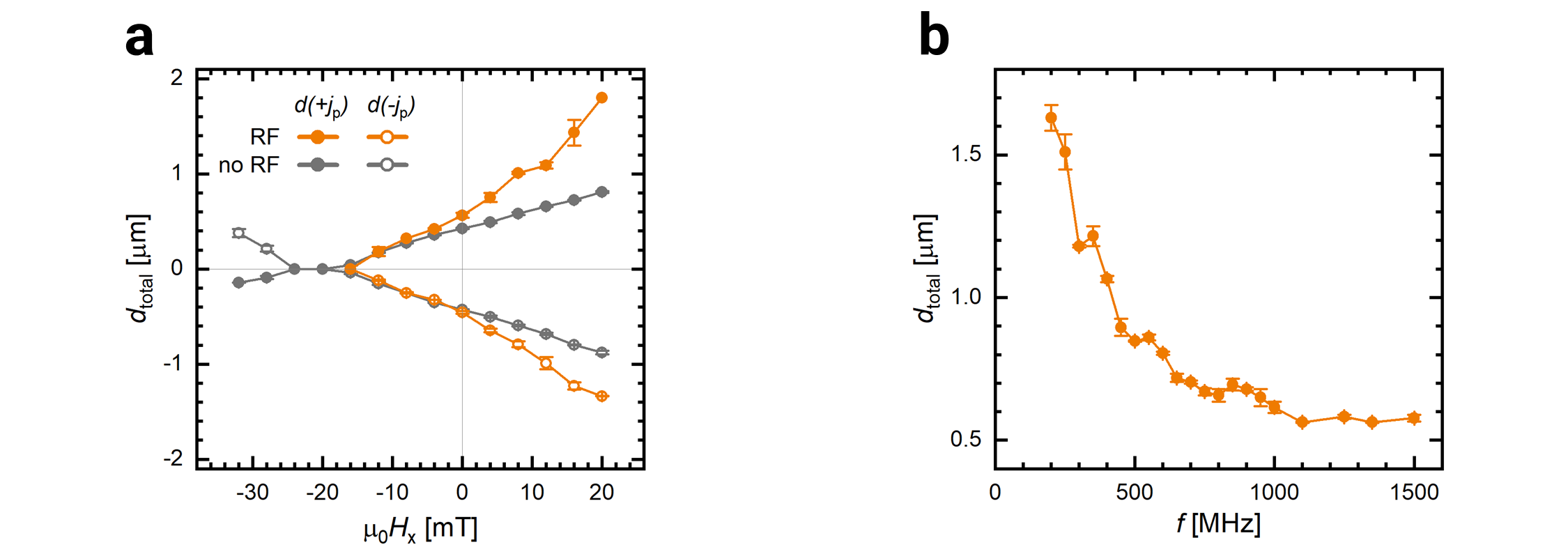}
    \caption{\textbf{Field and frequency dependence of sustained domain wall motion.} 
    In both graphs, current pulses with \jp{} $=136$ \MAcm{} and \pulseT{} $=5$ ns and RF currents with a constant output power of \mbox{\PowOutNew{} $=-3$ dBm} are applied.
    \textbf{(a)} An RF current with $f=250$ MHz and a duration \tRF{} $=500$ ns is used.
    \textbf{(b)} A longitudinal magnetic field \muHx{} $=8$ mT is applied to a down-up DW. The applied RF current has a duration of \tRF{} $=1.0$ \MuSec{}.
    }
    \label{fig-SI: sus-mot Hx frequency}
\end{figure}

\begin{figure}
    \centering
    \includegraphics[width=\linewidth]{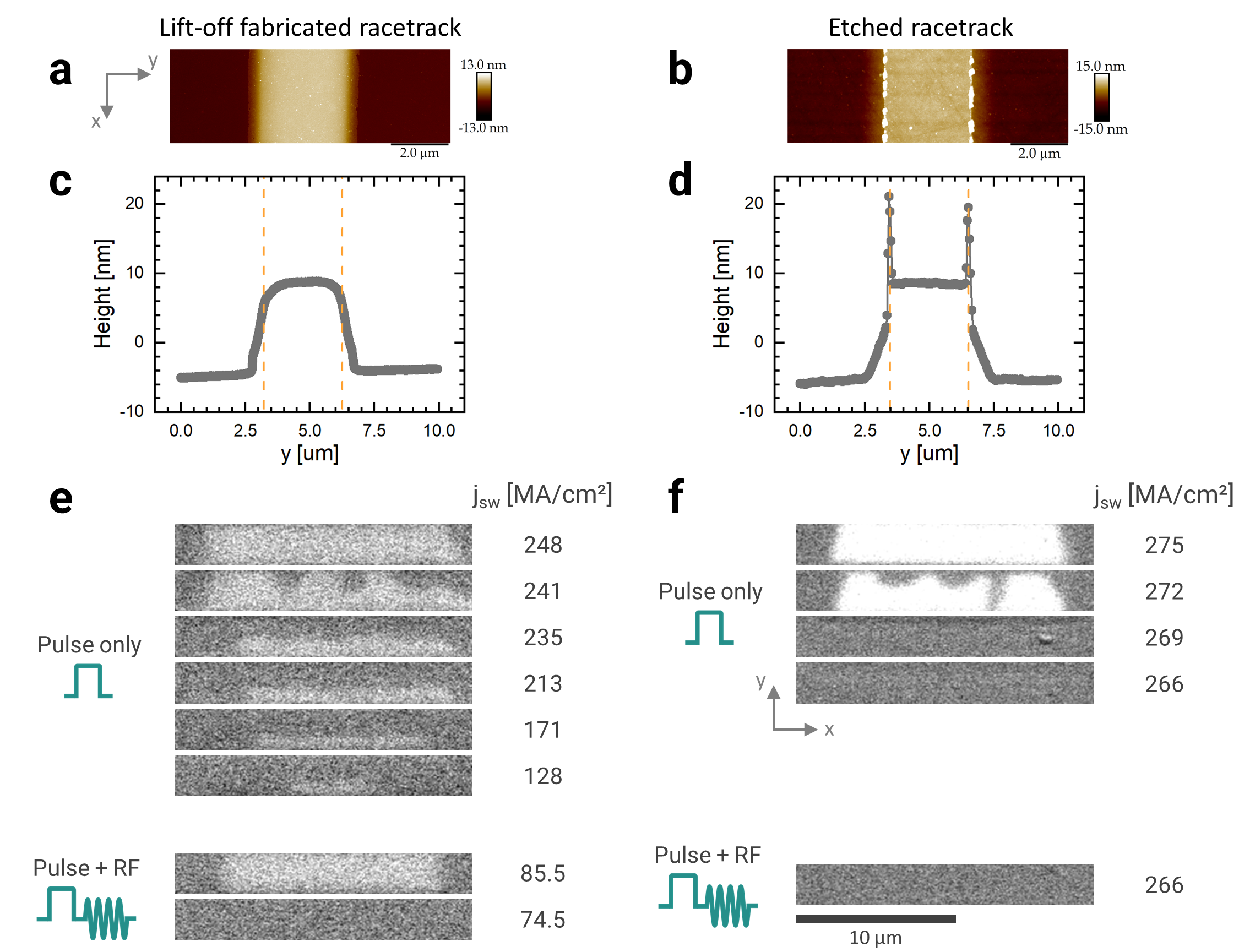}
    \caption{\textbf{Comparison of different fabrication routines with respect to racetrack cross-section and switching characteristics.} 
    The data in the left column investigates a typical racetrack device used in this paper, being fabricated by a lift-off technique. In contrast, the right column demonstrates the properties of a reference device that is fabricated using Ar$^+$ ion etching. 
    \textbf{(a,b)} The topography of the racetrack is scanned via atomic force microscopy (AFM). 
    The height profile along a y-cut in \textbf{(c,d)} shows that the typical racetrack (left) gradually flattens and shows a smooth variation of its height. In contrast, the etched racetrack (left) exhibits sharp, narrow spikes in its profile, followed by gradual flattening.
    \textbf{(e,f)} Up-to-down switching of the racetrack in a longitudinal field \muHx{} $=+16$ mT using current pulses with \pulseT{} $=10$ ns. Both the single pulse (top) and a pulse-RF combination (bottom) are used for switching. An RF current with $f=100$ MHz, \tRF{} $=5$ \MuSec{} and \PowOutNew{} $=-2.1$ dBm is used. The lift-off fabricated racetrack exhibits a lower threshold for full switching that the etched one, but it also exhibits a broad range of pulse current densities where edge switching is observed. No edge switching is observed for the etched racetrack. Moreover, for the lift-off fabricated racetrack, a pulse-RF combination leads to full switching even at current densities where no pulse-induced edge switching could be resolved experimentally anymore. No such RF-assisted switching arises for the etched racetrack, even at considerably higher currents.
    }
    \label{fig-SI: AFM switching}
\end{figure}

\begin{figure}
    \centering
    \includegraphics[width=\linewidth]{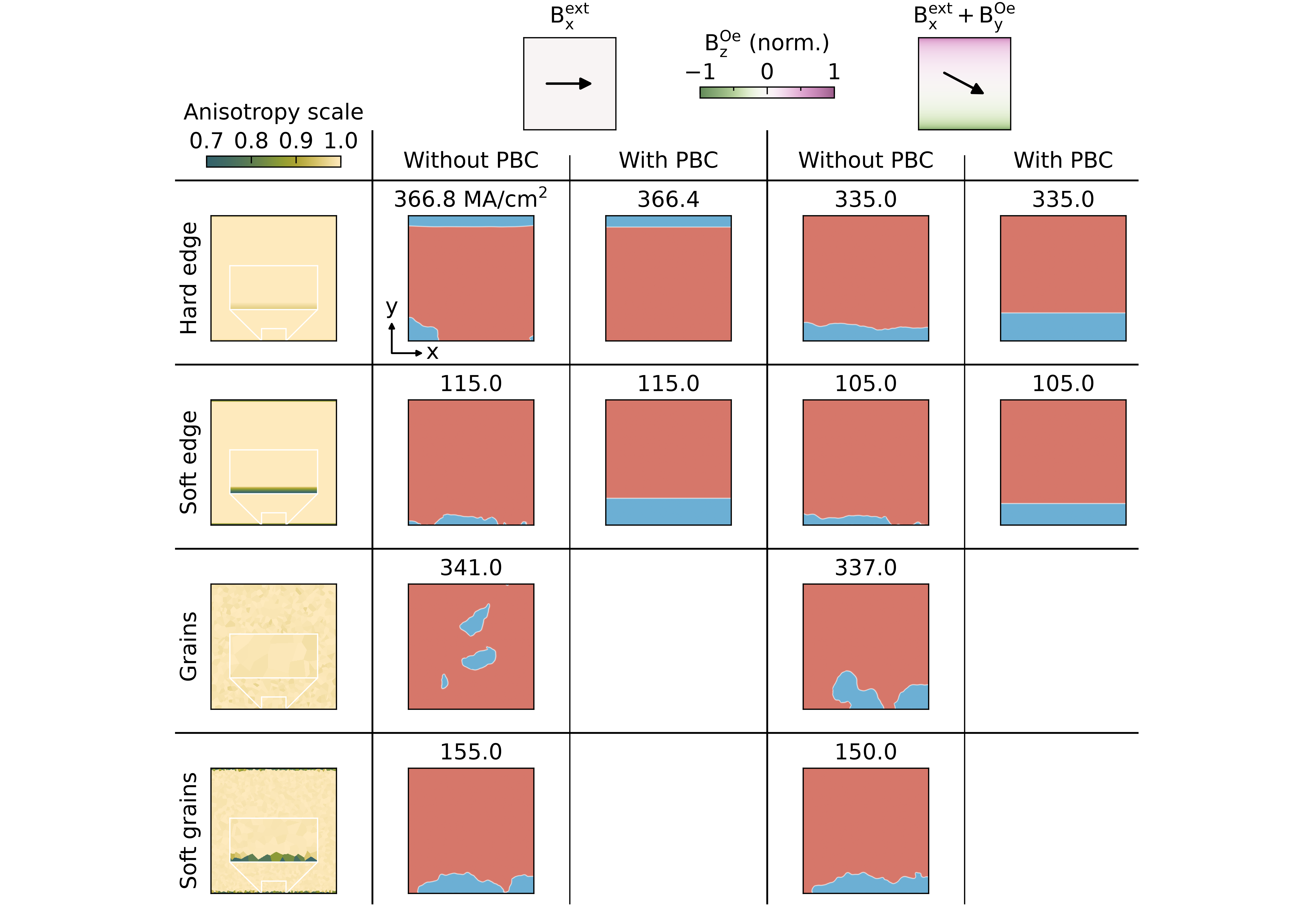}
    \caption{\textbf{Asymmetry of edge domain nucleation as a function of edge softening, Oersted field, and boundary conditions.} 
    Rows compare different edge-softening approaches, major columns compare simulations without and with the current-induced Oersted field $\mathrm{B_{y,z}^{Oe}}$, and minor columns compare simulations with and without periodic boundary conditions (PBCs). The simulation box is initialized in the up-magnetized state (red). Snapshots are taken approximately 3 ns after the end of a 2 ns current pulse along the positive $x$ direction. The current density above each snapshot is chosen near the nucleation threshold of a reverse domain (blue).
    The upper two rows employ a uniform film with symmetric edge softening, whereas the lower two rows use a Voronoi grain structure. Simulations with PBCs along x remove the demagnetizing fields associated with the left and right edges, thereby emulating an infinitely long racetrack. With the Oersted field \mbox{(columns 4 and 5)}, the total applied field is given by $\mathrm{B_x^{ext}}$ and $\mathrm{B_{y,z}^{Oe}}$.
    Four main trends emerge. First, without pronounced edge softening or Oersted field, nucleation occurs preferentially at the upper edge, whereas magnetically softened edges reverse this asymmetry, resulting in the experimentally observed lower-edge nucleation while substantially reducing the threshold current density. Second, inclusion of the Oersted field produces a similar reversal of the edge-switching asymmetry. Third, introducing a grainy film morphology does not qualitatively alter the trends induced by edge softening or the Oersted field. Finally, PBCs suppress edge-induced perturbations of the injected DW, resulting in a straight wall, whereas open boundaries produce ripples, bubbles, and kinks.
    Overall, the simulations reproduce the experimentally observed asymmetric lower-edge nucleation for positive current and field directions. Moreover, the nucleation thresholds obtained for magnetically softened edges agree well with the experimentally observed current density range  (Supplementary Fig. \ref{fig-SI: AFM switching}).
    }
    \label{fig-SI: Mumax switching asymmetry}
\end{figure}

\begin{figure}
    \centering
    \includegraphics[width=\linewidth]{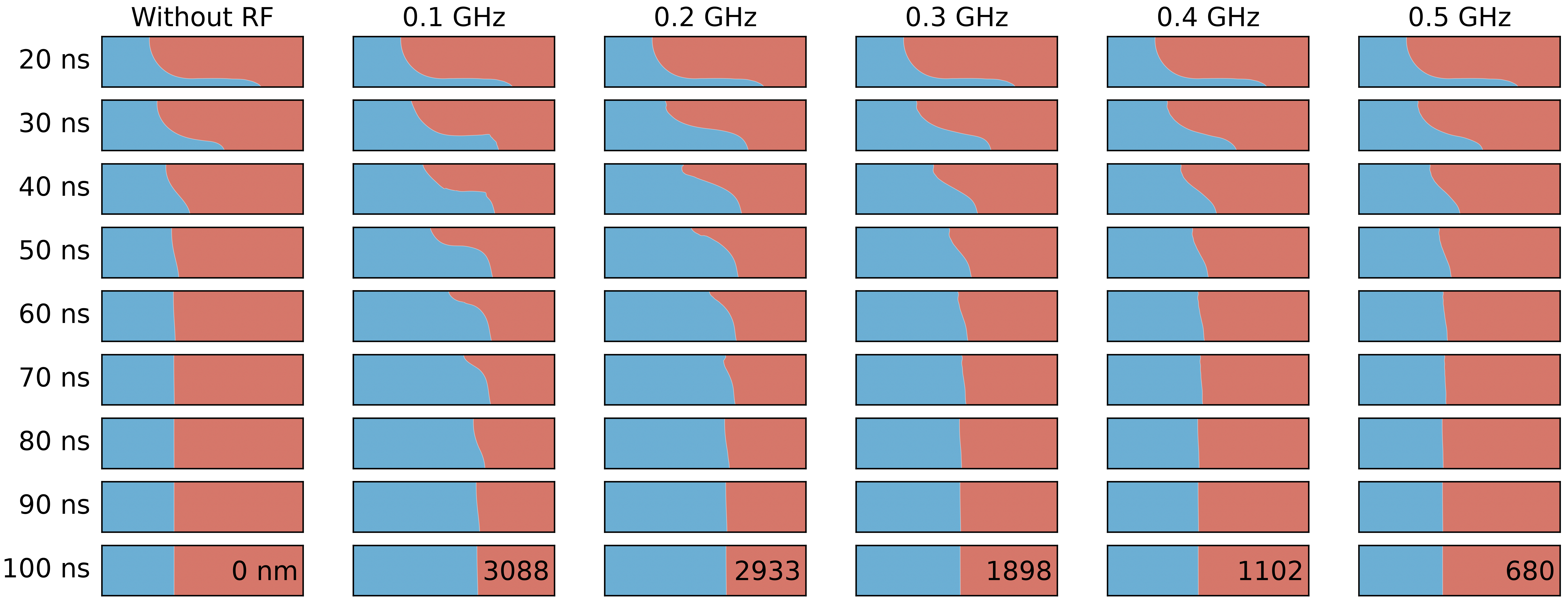}
    \caption{
    \textbf{Frequency dependence of sustained domain wall motion.} 
    Using the same simulation setup as in Fig. \ref{fig_main: sus mot explain}e and starting from the same post-pulse magnetization state at $20$ ns, an RF current is applied for $50$ ns and its frequency is varied from \mbox{$0.1$ to $0.5$ GHz}. The resulting sustained DW motion is compared with the DW relaxation in the absence of an RF excitation (left column). Columns correspond to different RF frequencies, while rows show the magnetization state at 10 ns intervals. For the shown frequencies, the snapshots in each row correspond to the same RF phase across all columns. The bottom row indicates the final displacement of the vertical DW and the values represent the additional displacement relative to the DW that relaxed without RF excitation, expressed in nano-meters. For the RF and field parameters considered here, $f_\mathrm{RF}=0.1$ GHz produces the largest additional DW displacement and therefore the most effective sustained DW motion. This frequency dependence is consistent with the edge domain growth rate analysis (see Supplementary Fig. \ref{fig-SI: Mumax sus-mot mechanism}).
    }
    \label{fig-SI: Mumax sus-mot freq-dep}
\end{figure}

\begin{figure}
    \centering
    \includegraphics[width=\linewidth]{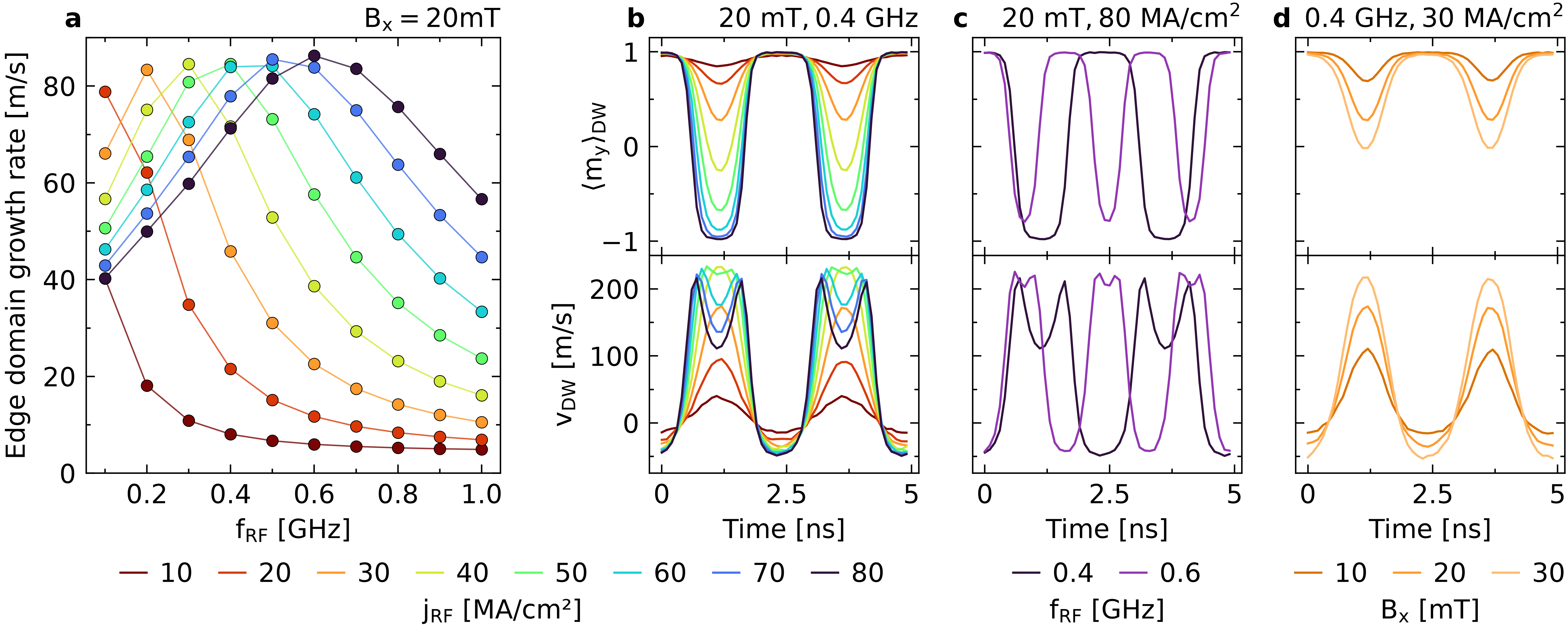}
    \caption{
    \textbf{Parameter dependence and time evolution of RF-driven edge-domain expansion.}
    To better understand the experimentally observed and numerically reproduced frequency dependence of sustained DW motion (see Supplementary Figs. \ref{fig-SI: sus-mot Hx frequency}b and \ref{fig-SI: Mumax sus-mot freq-dep}), the RF-driven edge domain expansion was studied in more depth. Similar to the simulation setup in Fig. \ref{fig_main: sus mot explain}d, a horizontal DW was initialized close to the bottom edge of the racetrack while periodic boundary conditions along the x-direction were applied. 
    \textbf{(a)} For each RF current density $\mathrm{j_{RF}}$, the net edge domain growth rate reaches a maximum at a characteristic frequency $\mathrm{f_{RF}}$. \textbf{(b)} to \textbf{(d)} show the time evolution of the mean $y$-component of the normalized magnetization within the DW, \mbox{$\langle \mathrm{m_y} \rangle _\mathrm{DW}$ (top)}, and the vertical DW velocity, $\mathrm{v_{DW}}$ (bottom), at high temporal resolution for selected parameter sets. Both the average and peak DW velocity are maximal when the DW magnetization nearly reverses from $+y$ to $-y$ during the positive current half-cycle. For positive current, this reversal results from the competition between the DMI-stabilized DW configuration and the SOT-induced rotation of the DW magnetization. At insufficient RF-drive - i.e. low $\mathrm{j_{RF}}$, high $\mathrm{f_{RF}}$ and/or low $\mathrm{B_x}$ - the DW remains close to equilibrium, resulting in small displacements. At excessive RF drive - i.e. high $\mathrm{j_{RF}}$, low $\mathrm{f_{RF}}$ and/or high $\mathrm{B_x}$ - the DW reverses its $y$-component from $+y$ to $-y$ within a shorter time. As a result, the DW spends a larger fraction of the current half-cycle aligned close to $-y$, where its instantaneous velocity is reduced, leading to a lower mean velocity. During the negative current half-cycle, the DW moves in the opposite direction with a significantly lower velocity throughout all parameter sets. In the reversed direction, high $\mathrm{j_{RF}}$ and high $\mathrm{B_x}$ lead to a faster velocity. Thus, the RF-driven edge DW expansion depends on the interplay of RF-current amplitude, RF frequency, and applied longitudinal field amplitude.
    \mbox{\textbf{(b)} $\mathrm{j_{RF}}$ is} varied at fixed $\mathrm{B_x}$ and $f\mathrm{_{RF}}$.
    \textbf{(c)} $\mathrm{f_{RF}}$ is varied at fixed $\mathrm{B_x}$ and $\mathrm{j_{RF}}$.
    \textbf{(d)} $\mathrm{B_x}$ is varied at fixed $\mathrm{f_{RF}}$ and $\mathrm{j_{RF}}$.
    }
    \label{fig-SI: Mumax sus-mot mechanism}
\end{figure}

\begin{figure}[]
    \centering
    \includegraphics[width=\linewidth]{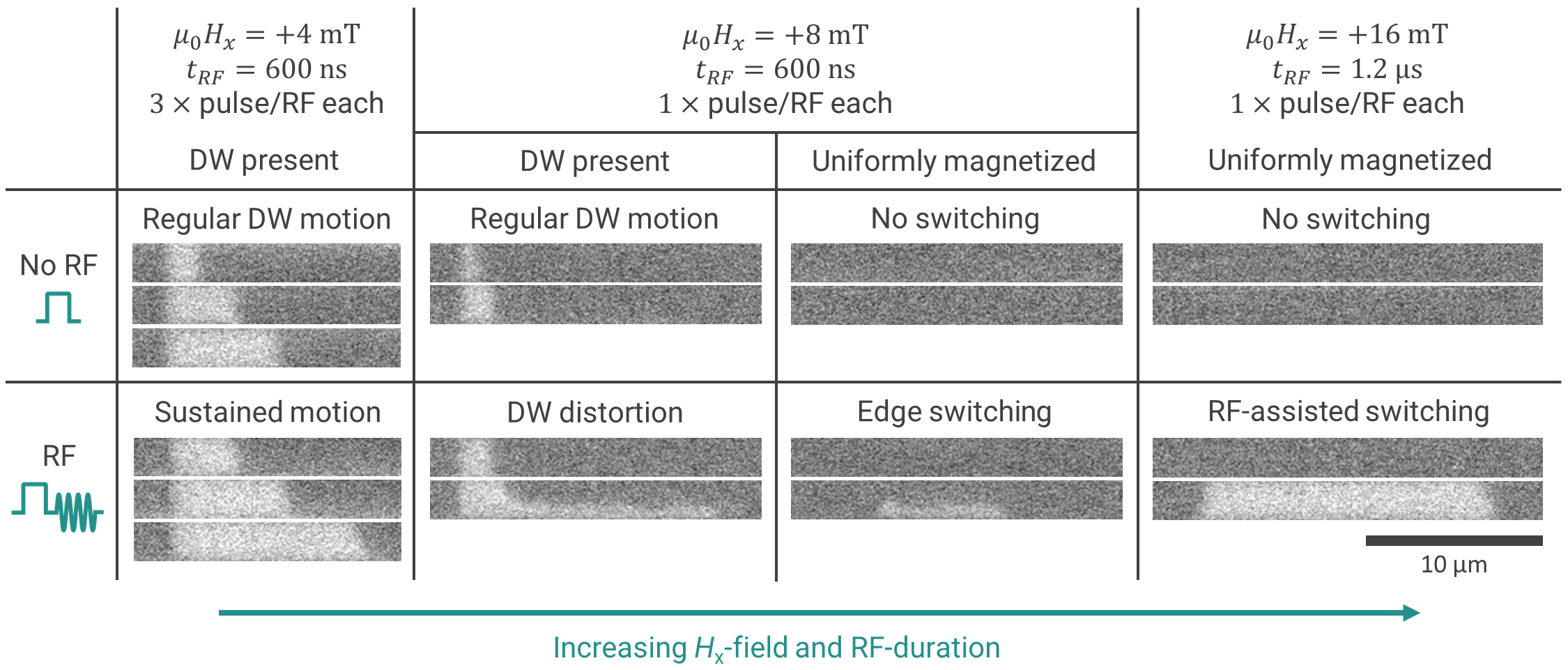}    
    \caption{\textbf{Transition from sustained domain wall motion to magnetic switching.}
    Kerr microscopy snapshots showing the evolution from sustained DW motion (left) through an intermediate regime characterized by DW distortion and edge switching (center) to RF-assisted switching (right). Increasing \Hx{} and \tRF{} progressively transforms sustained propagation into complete magnetic reversal. Without RF excitation (top row), neither DW distortion nor switching is observed. For the two rightmost columns, the upper gray images show the initial uniformly magnetized state, whereas the lower images show the magnetic configuration after application of a single pulse or pulse–RF sequence. Pulses with \jp{} $=214$ \MAcm{}, \pulseT{} $=7.5$ ns and RF-bursts with \mbox{$f=100$ MHz}, \PowOutNew{} $= -2.1$ dBm are used.
    }
    \label{fig-SI: evolution sus-mot switching}
\end{figure}

\begin{figure}
    \centering
    \includegraphics[width=\linewidth]{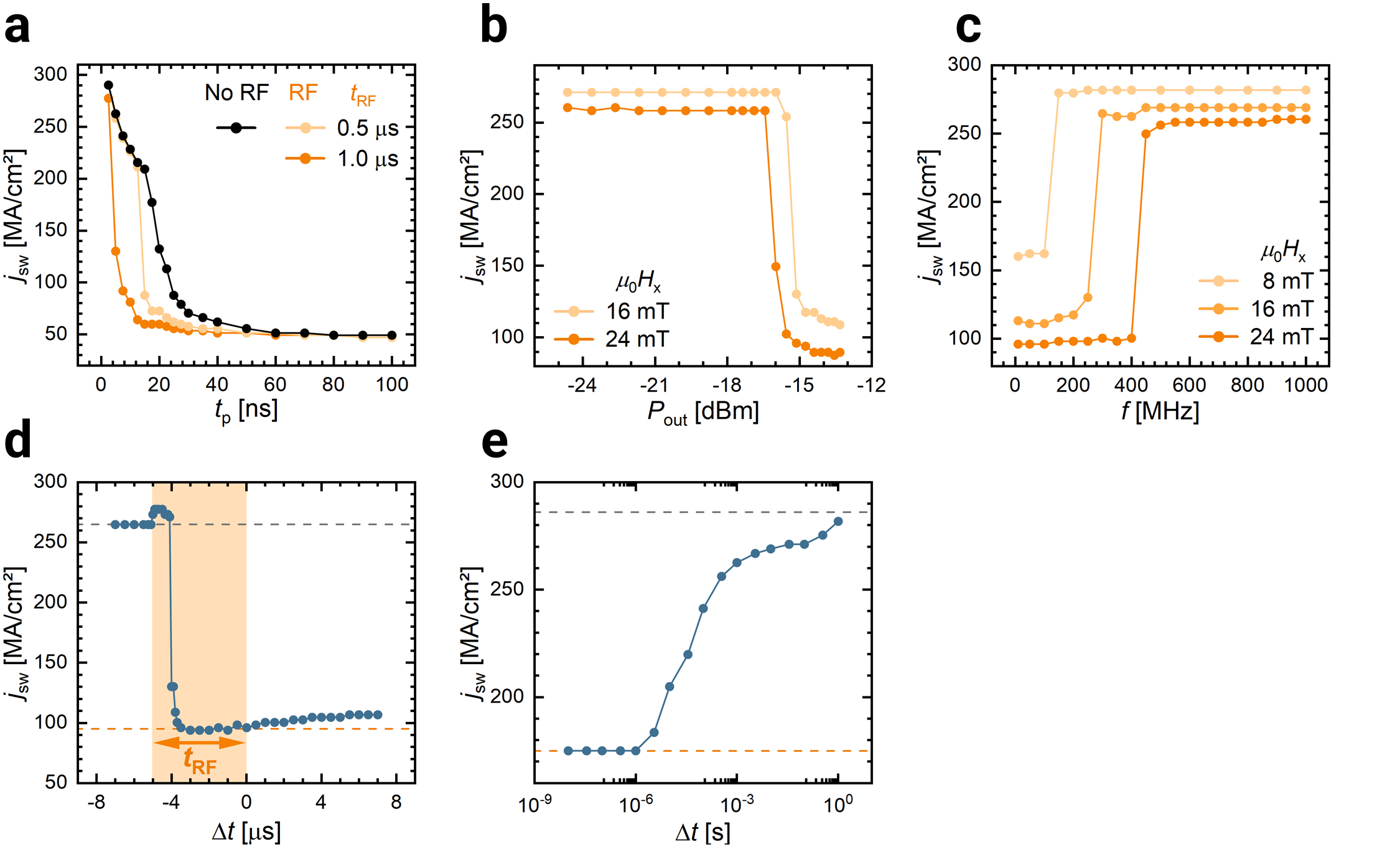}
    \caption{\textbf{Additional parameter dependences of RF-assisted magnetic switching.} 
    Dependences of the switching current density \jsw{} on the current pulse width \textbf{(a)}, transmitted RF power \textbf{(b)}, RF frequency \textbf{(c)}, and the relative timing between current pulse and RF burst \textbf{(d,e)}. Unless stated otherwise, a longitudinal field \mbox{\muHx{} $=24$ mT} is applied and a current pulse with \pulseT{} $=5$ ns arrives immediately before the RF burst \mbox{($\Delta t \approx 10$ ns)}. The RF frequency is set to $f=100$ MHz and the transmitting power amounts to \PowOutNew{} $=-3$ dBm. Exceptions are an RF frequency of $f=250$ MHz in \textbf{(b)} and a field of \muHx{} $=8$ mT in \textbf{(e)}.
    }
    \label{fig-SI: RF switching params}
\end{figure}

\begin{figure}[h!]
    \centering
    \includegraphics[width=\linewidth]{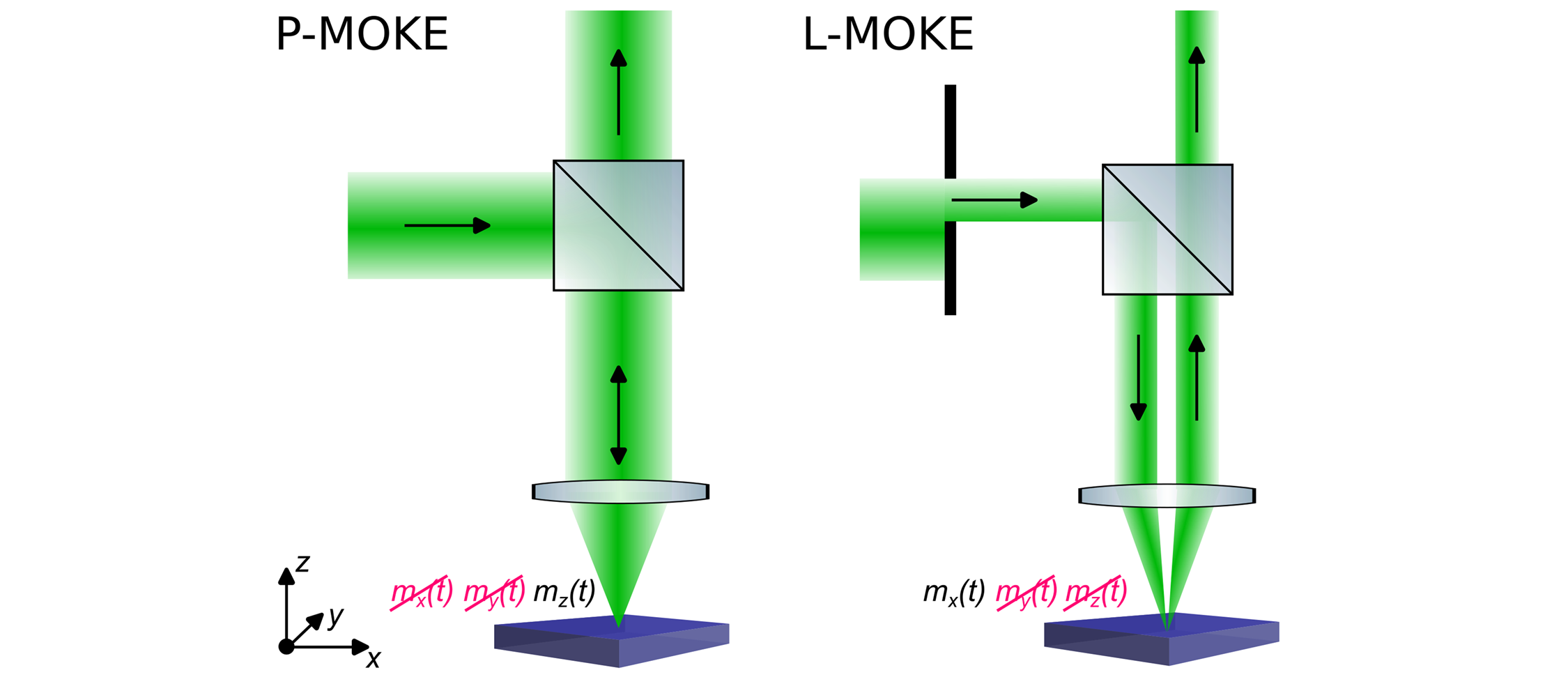}
    \caption{\textbf{Schematic of the illumination paths during the P-MOKE and L-MOKE configuration during SNS-MOKE.}}
    \label{fig-SI: Illumination_path}
\end{figure}

\begin{figure}[h!]
    \centering
    \includegraphics[width=0.5\linewidth]{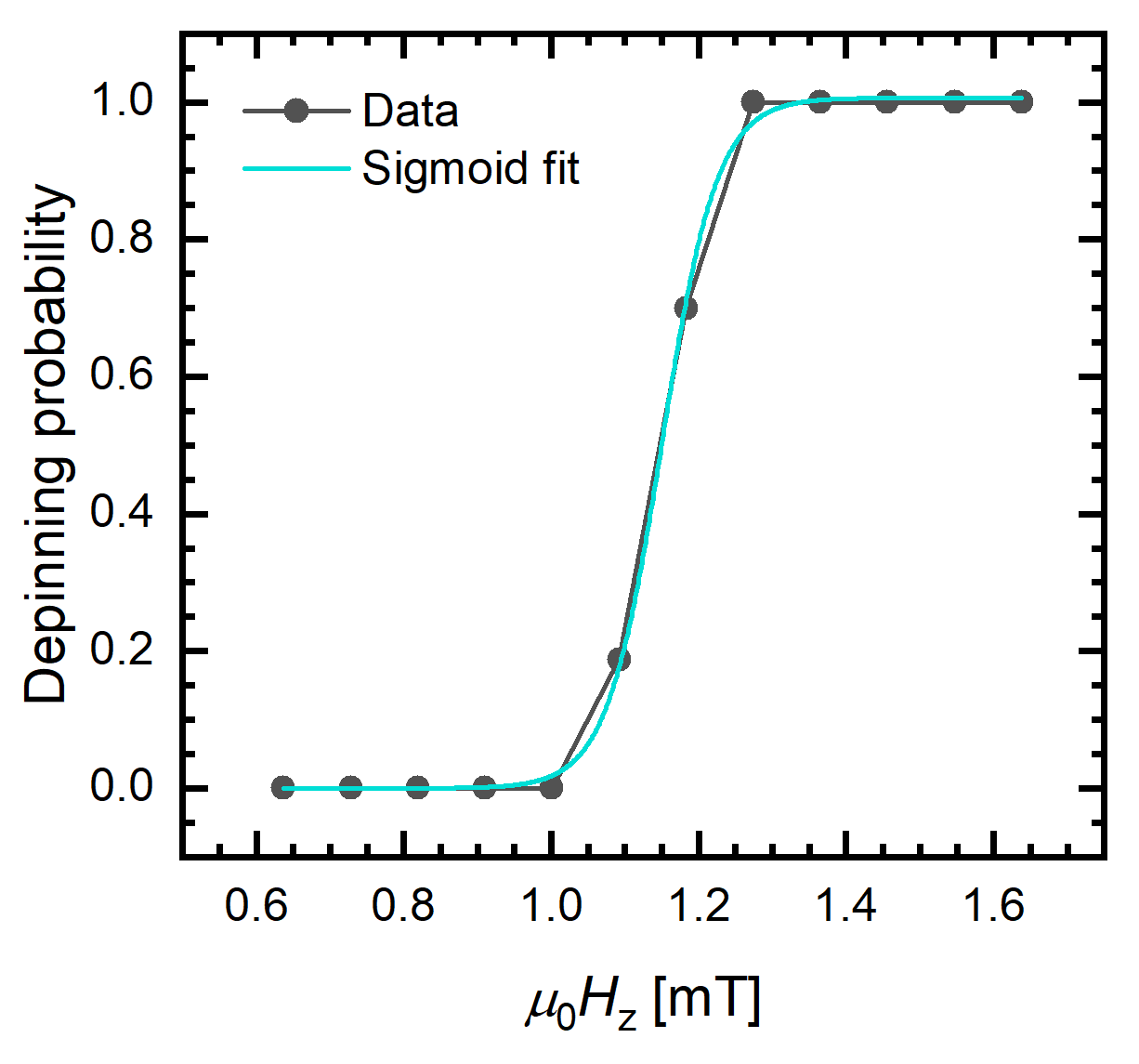}
    \caption{\textbf{Determination of the out-of-plane depinning field.} An isolated DW is initialized in a region with weak extrinsic pinning. An out-of-plane magnetic field is applied for 1 s and the probability of DW displacement is determined from the fraction of trials in which DW motion is observed. The DW is reinitialized at the same position before each trial. A sigmoidal fit to the measured probabilities is used to determine the depinning field, defined as the field corresponding to a 50\% probability of DW displacement.}
    \label{fig-SI: OOP depinning field}
\end{figure}

\begin{figure}[h!]
    \centering
    \includegraphics[width=0.8\linewidth]{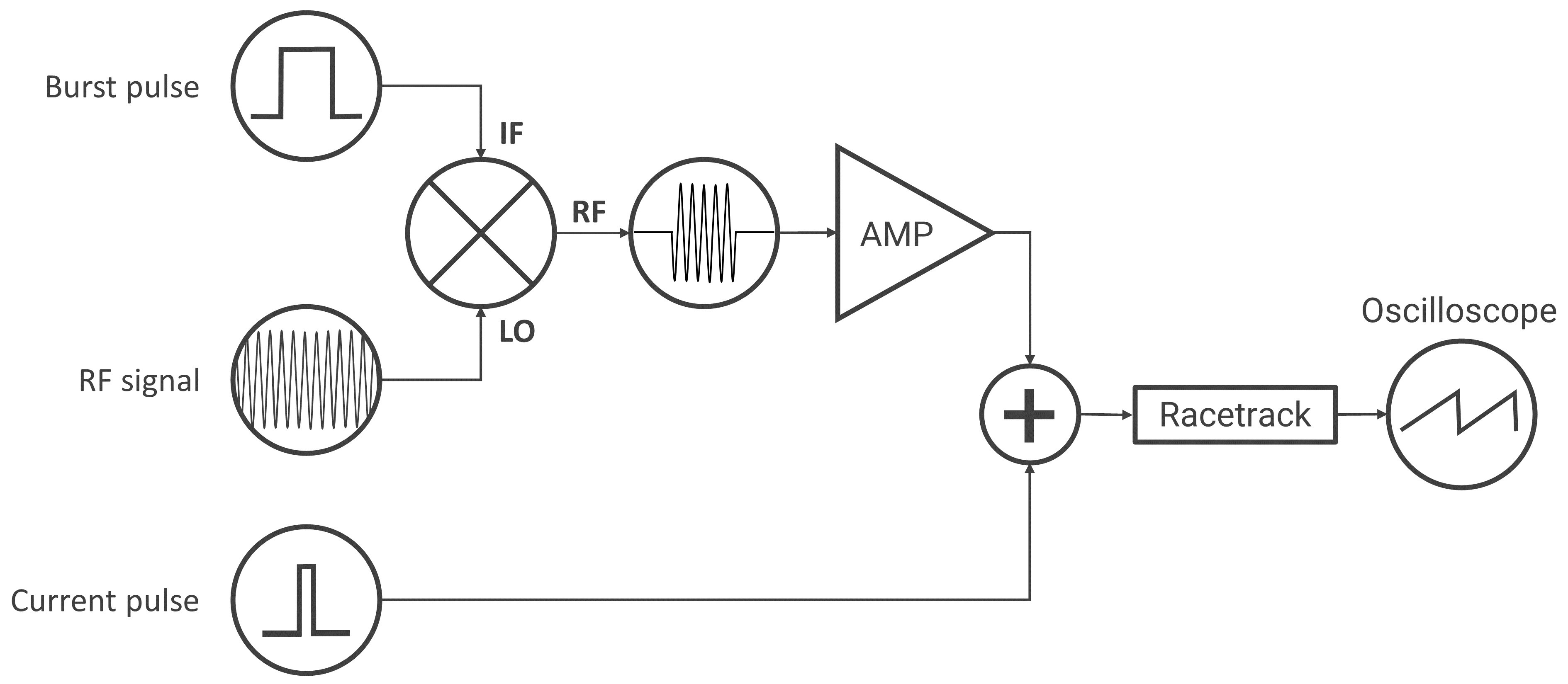}
    \caption{\textbf{Schematic of electronic circuitry setup for simultaneous pulsed current and RF current injection.}}
    \label{fig-SI: circuitry}
\end{figure}

\end{document}